\documentclass[article,12pt]{article} 

\usepackage[utf8]{inputenc}

\usepackage{amscd,amsmath,amssymb,amsfonts,latexsym,mathrsfs,amsthm}
\usepackage{graphicx} 
\usepackage{setspace} 
\usepackage{graphics,epsfig}
\usepackage{sectsty}
\usepackage[authoryear]{natbib}
\usepackage[english]{babel}
\usepackage{amsmath}
\usepackage{amsfonts}
\usepackage{amssymb,amsthm}
\usepackage{bm}
\usepackage{mathrsfs}
\usepackage{subfigure}
\usepackage[usenames]{color}
\usepackage{rotating}
\usepackage{multirow}

\usepackage{url}

\newtheorem{rem}{{\bf Remark}}

\usepackage{flushend}
\usepackage{verbatim}
\usepackage{tabularx}
\usepackage{multirow} 
\usepackage{arydshln}
\usepackage{subfigure}

\UseRawInputEncoding

\title{An automatic counting algorithm for the quantification and uncertainty analysis of the number of microglial cells trainable in small and heterogeneous datasets} 
\author{L. Martino$^\diamond$\footnote{Corresponding author: luca.martino@unict.it}, M. M. Garcia$^\dagger$, P. S. Paradas$^\dagger$, E. Curbelo$^\ddagger$, \\
{\small$^\diamond$ Universit{\'a} degli studi di Catania, Catania, Italy.}\\
{\footnotesize$^\dagger$ Area of Pharmacology, Nutrition and Bromatology, Dep. of Basic Health Sciences,}\\ 
{\footnotesize Faculty of Health Sciences, Universidad Rey Juan Carlos (URJC),}\\
 {\footnotesize Unidad Asociada de I+D+i al Instituto de Quimica Medica (IQM), CSIC-URJC.} \\
{\small$^\ddagger$ Universidad Carlos III de Madrid, Madrid, Spain.}\\
}

\date{}
 
\begin{document}

\maketitle

\thispagestyle{empty}

\begin{abstract}
Counting immunopositive cells on biological tissues generally requires either manual annotation or (when available) automatic rough systems, for scanning signal surface and intensity in whole slide imaging. In this work, we tackle the problem of counting microglial cells in lumbar spinal cord cross-sections of rats by omitting cell detection and focusing only on the counting task. Manual cell counting is, however, a time-consuming task and additionally entails extensive personnel training. The classic automatic color-based methods roughly inform about the total labeled area and intensity (protein quantification) but do not specifically provide information on cell number. Since the images to be analyzed have a high resolution but a  huge amount of pixels contain just noise or artifacts, we first perform a pre-processing generating several filtered images {(providing a tailored, efficient feature extraction)}. Then, we design an automatic kernel counter that is a non-parametric and non-linear method. The proposed scheme can be easily trained in small datasets since, in its basic version, it relies only on one hyper-parameter. However, being non-parametric and non-linear, the proposed algorithm is flexible enough to express all the information contained in rich and heterogeneous datasets as well (providing the maximum overfit if required). 
Furthermore, the proposed kernel counter also provides uncertainty estimation of the given prediction,  and can directly tackle the case of receiving several expert opinions over the same image.  Different numerical experiments with artificial and real datasets show very promising results. Related Matlab code is also provided. 
\end{abstract}

\section{Introduction}
\label{sec-intro}

\begin{figure}[h!]
    \centering
    \centerline{
 %  \subfigure[\label{FigEntera}]{\includegraphics[width=15cm]{5.png}}
    %\includegraphics[width=15cm]{5.png}
\includegraphics[width=15cm]{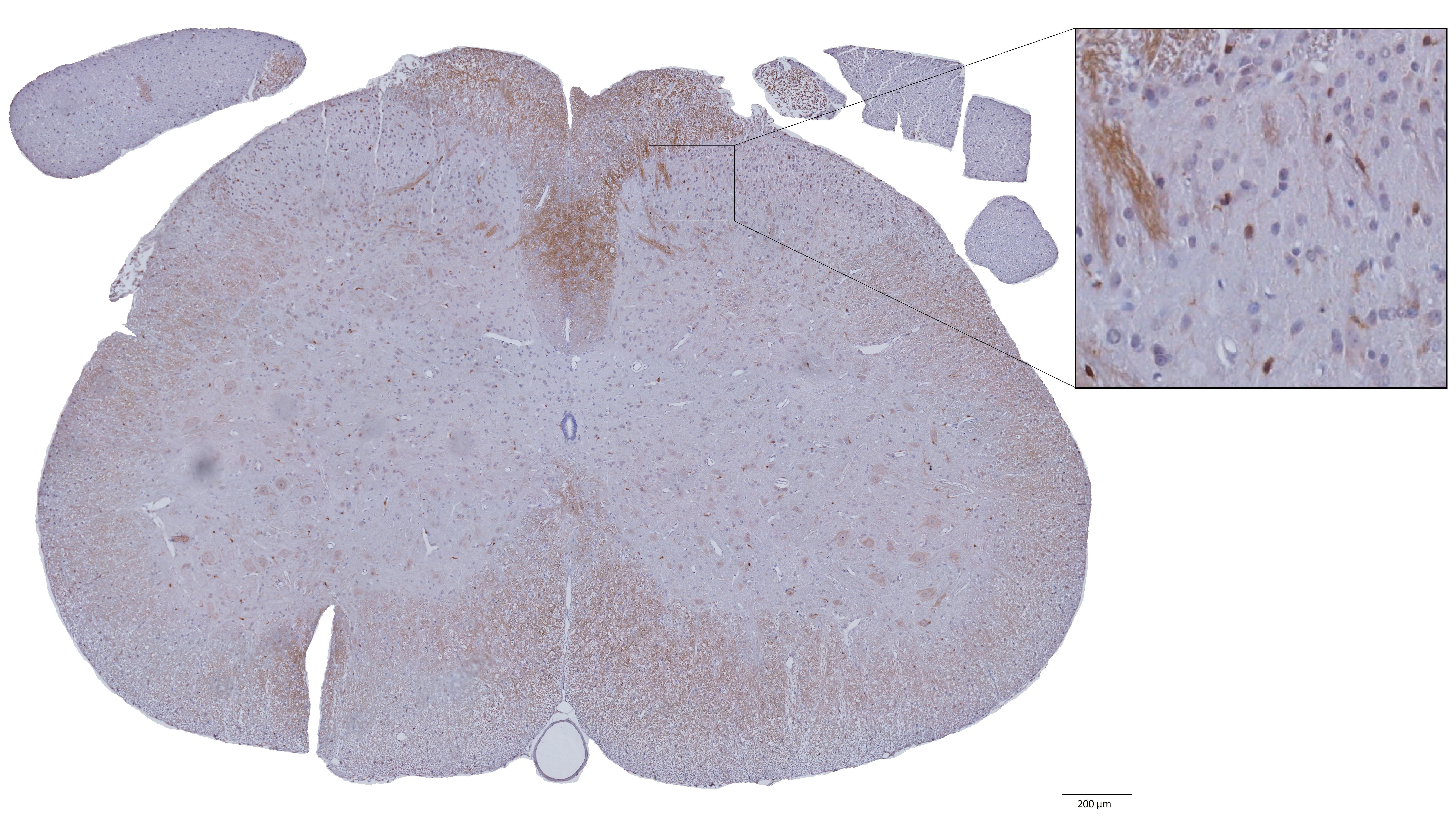}

    }
    \caption{{\footnotesize %{\bf(a)} 
    Example of complete image (spinal cord cross-section).  The zoom frame indicates part of the ipsilateral dorsal horn with microglial cells (scale = 200$\mu m$). %{\bf(b)} A portion of an image where some microglial cells have been highlighted with red pixels. 
     } }
   % \label{FigGenMicro}
   \label{FigEntera}
\end{figure}

\begin{figure}[h!]
    \centering
    
  \centerline{ \includegraphics[width=12cm]{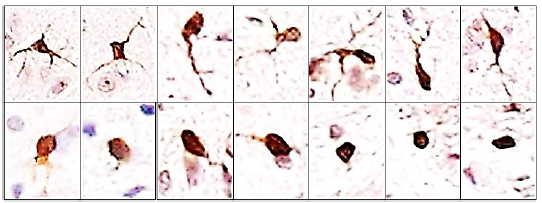}
   }
%       \centerline{
%      \includegraphics[width=12cm]{9_X_v2.png}
%    }
    \caption{{\footnotesize Examples of microglial cells (up: ramified; down: amoeboid). Microglia is always
represented by a brownish tinction, whereas blue colored cells usually correspond to the
nuclei of neurons.} }
    \label{FigGenMicro2}
\end{figure}

The problem of counting objects in images or video frames is still one of the relevant tasks that many biomedical applications face \cite{lempitsky2010learning,Ongena2010}.
Cell counting in images is yet a relevant but unpolished issue for many applications \cite{Bradesi10,Kongsui2014}. Cell counting on tissue sections is typically a tedious and time-consuming task, and additionally entails extensive personnel supervision and training.
The cells called microglia constitute one major group of glial cells,  located throughout the brain and spinal cord of the central nervous system. They are considered resident immune cells within the central nervous system and increase in size and number upon activation \cite{trang_beggs_salter_2011}. Moreover, they have been suggested to be responsible for initiating altered synaptic and firing activity in neurological disorders, including chronic neurodegenerative diseases (e.g., Alzheimer's and Parkinson's disease) and chronic pain \citep{MINGHETTI2005251}. However, one major concern when analyzing the microglial cells is their identification and counting
\cite{PlosOneAutoMicro,Green2022,Suleymanova2023}.
\newline
\newline
{ Microglial cells exhibit a variety of shapes, sizes and functions depending on their activation
state.  They show fine, highly ramified processes that are thin and often intertwined.
These features complicate both manual and automatic tracing methods, especially in noisy or low-contrast microscopy images \cite{Heindl2018-cg,Kwok21,McKay2007-cg,Mecha2016-jm}. 
Therefore, often methods for quantifying cells imply undeniable bias and suffer several limitations regarding usability, sensitivity, and robustness.
Thus, manual counting remains the benchmark technique in many studies. However, this approach is laborious, monotonous and susceptible to both intra- and inter-operator variability \cite{degracia2015automatic,gallego2016automatic,ly2021c3vfc,reddaway2023microglial}.}
\newline
\newline
{ Manual counting is the standard when dealing with this counting problem \cite{gallego2016automatic}. Software solutions (as ImageJ   \cite{schneider2012nih}) based on image filtering can help in the automation of the process \cite{de2015automatic, khakpour2022manual}. 
The use of these softwares exhibit a quite high correlation with human expert results but the process is still not completely automatic based on specific threshold colors and other user decisions. Often, they also require additional information, such as the scale in micrometers per pixel etc. 
Deep learning methods such as convolutional neural networks (CNN) have also been applied.
They learn patterns {\it directly from the high-resolution images} \cite{anwer2023comparison, mohle2021development, suleymanova2018deep}. These approaches show significant improvement in time once the model is trained, but the training process is still long and based on the previous detection of the cells. Hence, the human expert  must label all the microglial pixels of the image (instead of just counting cells) for the dataset creation.
 Furthermore, some additional restrictions persist in the dataset creation, including the necessity for all images to be of uniform dimensions (the same size). 
 Last but not least, providing uncertainty estimates in the results and addressing label uncertainty remain challenging tasks that are also tackled in this work. }
\newline 
\newline 
 In this work, we focus only on the simplified task of just counting  problem of the microglial cells {(when a 3,3-Diaminobenzidine (DAB) immunohistochemistry staining is performed)}, skipping to necessarily detect them. We consider the microglial cells in micrographs of lumbar spinal cord cross-sections of rats (see Figure \ref{FigGenMicro2}). They exhibit a variety of shapes, sizes, and functions depending on their activation state  \cite{Green2022,Wittekindt2022-in}. 
The problem of manual counting microglial cells in immunohistochemistry is typically a time-consuming task, often requiring a budget for hiring personnel devoted to this task and  additionally entails extensive personnel training \citep{Bradesi10,GARCIA2022112986,trang_beggs_salter_2011}. Indeed, computing immunopositive cells on biological tissues generally requires the ability to detect the cells and manual annotation  \citep{GARCIA2022112986,Morelli21}.  
 This cell quantification is not possible with other immunoblotting techniques like ELISA (enzyme-linked immunosorbent assay) or western blot for tissue samples to be homogenized \cite{Kwok21}.
 { Several specific issues and features of the tackled problem (i.e., counting microglial cells) should be taken into account}:  
%Microglial cells are very small and stained a distinguishable dark brown color. On the contrary, size and shape depend on both the cutting plane and activation state [2, 21]. 
\begin{itemize}
%\newline 
\item[{\bf (a)}] Microglial cells are very small and stained a distinguishable dark brown color. On the contrary, size and shape depend on both the cutting plane and activation state \citep{Bradesi10,trang_beggs_salter_2011}. This is why using shape information to count (and/or detect) them can be misleading for the learning algorithm.
%%%%%%%%%%%%
\item[{\bf (b)}] The images to be analyzed have a high resolution,
where a huge amount of pixels are just noise or artifacts. The number of pixels forming the microglial cells is extremely small with respect to the number of pixels that are not contained in a cell. The difference in order of magnitude, in terms of number of pixels, is approximately $10^4$ in favor of non-microglial objects (on average).  For instance, Figure \ref{FigGenMicro2} depicts just very tiny portions of an entire image, whereas an entire image is given in Figure \ref{FigEntera}. Namely, most of the input signals in our problem represent virtually ``noise'', and hence contain useless and/or misleading information.%%%%%%%%%%%%

%4. Given that dataset size depends on the inherent effort that demands manual processing, the number of analyzed images available in the dataset is small.
%5. Most importantly, the quality of images may be highly heterogenous within and among groups. Magnification, resolution, background counterstain, brightness, saturation, contrast and other color shifts might be present in the different image datasets. Hence the diversity in image quality from different laboratories may contribute to the need of a challenging algorithmic modeling.

\item[{\bf (c)}] The image dataset is created by a human expert after a long and tiresome  visual inspection.  This may unequivocally lead to computing errors. Furthermore, structural uncertainty is sometimes present for some cells. For these cases, even the expert is often not able to provide a clear decision (the expert could just provide an estimation of the uncertainty of being a microglial cell). 
%%%%%%%%%%%%

\item[{\bf (d)}]  The dataset size depends on the inherent effort that demands manual processing; hence, the number of stored and labeled images (denoted as $D$) available in the dataset is often small. The increase of the dataset depends mainly on costly human work (in terms of effort, time required, budget, etc). %%%%%%%%%%%% 
\item[{\bf (e)}]  Additionally, the quality of images in the dataset may be highly heterogeneous, depending on the imaging acquisition systems (jointly with human erroneous activities) in the laboratory. 
Differences in magnification, resolution, brightness, saturation, contrast, and other color shifts might be present in the stored images. 
Furthermore, the diversity grows if  different laboratories  share their images in order to increase the size of their datasets.
This heterogeneity contributes to the need for the design of suitable algorithms.
\end{itemize}
% \newline 
% \newline  
Hence, in this work, we design an automatic and adaptive counting method according to the requirements described above. Namely, we propose a flexible counting algorithm that can be easily trained with small datasets and is flexible enough to be able to express the diversity of images in the database. First of all, in order to address the issue of having a very low signal-to-noise ratio described in point {\bf (b)}, we perform a feature extraction filtering the images according to different color thresholds. This idea is based on the observation given in point {\bf (a)} above. After the filtering, we obtain binary filtered images with a much higher signal-to-noise ratio.
{ This first part, namely the tailored feature extraction, is described in Section \ref{Section3} (and denoted as P1)}. 
In the second stage, {denoted as P2 and described in Section \ref{SectionCore}}, we perform a kernel smoother to solve the regression problem, having as inputs the number of objects in those filtered images (counted by a clustering algorithm) and the expert's opinion as outputs.
The resulting algorithm, called {\it kernel counter} (KC), has the following characteristics:
 \begin{itemize}
\item KC is a {\it non-parametric} and {\it non-linear} method, i.e., its flexibility grows with the number $D$ of data/images in the dataset. Indeed, for every possible value of $D$, the method can work as an interpolator providing the perfect overfitting to the outputs, regardless of the diversity in the data.  
This characteristic then responds to the requirement {\bf (e)} described above. 
\item Moreover, in its basic version, the algorithm requires only the tuning of one hyper-parameter. Therefore, the learning task can be performed even in small datasets, fulfilling the condition {\bf (d)}.  Note that neural networks, or any other kind of parametric algorithms, can satisfy just one of the two requirements {\bf (e)} or {\bf (d)}, but not both together as the proposed KC does. 
\item The proposed KC method is also able to provide an uncertainty quantification of the given prediction. This  characteristic responds to the requirement {\bf (c)} above.
\item The proposed algorithm is a {\it counter}; indeed, the predictions/estimations are always non-negative. Thus, the predictions can easily converter into integers by rounding them.
\item The KC method can directly handle the case of receiving several expert opinions over the same image. Namely, different experts provide their count of the microglial cells for the same image. This is an isotopic multi-output scenario that can be directly handled by the KC \cite{MARTINOfusion} (more details are given in Section \ref{RobustenessSect}).
\end{itemize}
{ Furthermore, note that the proposed approach, based only in counting cells, facilitates substantially the dataset creation.}
We also discuss possible extensions and improvements to increase the robustness of the algorithm. Several additional theoretical and practical comments  are detailed in the appendices (e.g., regarding the choice of the threshold values).
 Different numerical experiments show the consistency of the proposed algorithm in different scenarios. We also test the KC method using images obtained by the Department of Basic Health Sciences, Faculty of Health Sciences of the Rey Juan Carlos University, Madrid, obtaining very promising results.  {Related Matlab code and the analyzed images are also provided at \url{http://www.lucamartino.altervista.org/PUBLIC_CODE_KC_microglia_2025.zip}.}

%y poder juntar database....algorithmo muy flexible...
 %(the only requirement is a white background in the image

%\begin{figure}[h!]
%    \centering
%
%       \centerline{
%   \includegraphics[width=10cm]{8.2.png}
%    }
%    \caption{{\footnotesize Graphical sketch of a microglial cell in 3 dimensions, and the different scenarios in which microglial cells can appear on a slice (2 dimensional representation). We also show a range of uncertainty level  ($0$ no  uncertainty, $1$ maximum  uncertainty) provided by the expert. } }
%    \label{FigGenMicro3}
%
%\end{figure}

%\newline
%\newline
%The work is structured as follows. %Some important features of the problem are highlighted in Section \ref{Section2}.
%  The main technical notations and ideas are introduced in Section \ref{Section3}. Section \ref{SectionCore} is devoted to the description of the main formulas of the proposed procedure, and to pose a notion of consistency. Section \ref{SectionGen} describes different extensions and improvements.  The choice of the thresholds is discussed in Section \ref{SectionChoice}. Some numerical tests and experiments are provided in Section \ref{SectionSimu}. We finish with some conclusions in Section \ref{SectionConc}.

{
\section{Key insights underlying the proposed approach}\label{KeySect}

The central idea is to focus solely on counting, rather than detecting, microglial cells. This choice has several important implications: for instance, it simplifies dataset creation and enables the design of more efficient computational algorithms. Moreover,
the  proposed (counting) algorithm  is divided into two main (and separated) parts, described below:
\begin{itemize}
    \item[{\bf P1}] The first part, described in Section \ref{Section3},  can be interpreted as a {\it tailored feature extraction} approach that  also represents a {\it significant  dimension reduction}: from any high-resolution image (or portion of it), we obtain a vector of non-negative values ${\bf r}_d$. 
In this way, we also achieve {\it effective denoising} 
by removing irrelevant input information and preserving the most informative components (of the input signal). This part has been specifically tailored (i.e., designed) to the microglial counting problem, taking into account the unique characteristics of the counting task that we aim to address: leveraging domain knowledge, the characteristics of the data, and database creation issues. It notably differs from the application of a generic black-box method typically used for feature extraction or dimensionality reduction, such as PCA, etc.
\item[{\bf P2}] For each labeled image in the database, we have an input vector  ${\bf r}_d$ associated with a value $N_d\geq 0$, provided by the expert (as the number of micriglial cell in the $d$-th image). Considering the input-output pairs $\{{\bf r}_d, N_d\}_{d}^{D}$, we now have a regression/prediction problem that could be theoretically solved by any regression method. However, we have several additional requirements: (a) predictions must always be non-negative; (b) the method should be non-linear and flexible; (c) it should remain easy to train even on small datasets; (d) 
it should have to possibility of increasing its flexibility by introducing additional hyperparameters, if needed; (e) it must provide uncertainty estimates for its predictions; and (f) ideally, it should also allow for the incorporation of expert uncertainty. Given these considerations, a kernel smoother emerges as a suitable candidate, as described in Sect. \ref{SectionCore}.
\end{itemize}
\noindent
{
The rest of the paper is devoted to describing P1 and P2 in details. We remark that any regression method could be theoretically applied in P2, although all the requirements must be satisfied.
}

}

%%%%%%%%%%%%%%%%%%%%%%%%%%%%%%%%%%%%%%
\section{Image filtering: extracting relevant information}\label{Section3}
\begin{figure}[h!]
    \centering
       \centerline{
   \includegraphics[width=15cm]{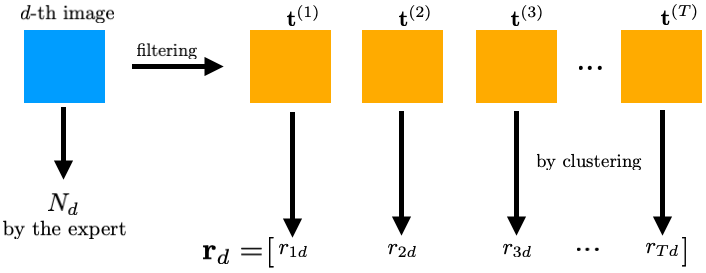}
    }
    \vspace{-0.3cm}
    \caption{{\footnotesize Graphical representation of the analysis performed for the feature extraction in each image. In each image filtered by ${\bf t}^{(k)}$, the total number of objects $r_{kd}$ is obtained by clustering. } }
    \label{FigSUPER_resumen_menos1}
\end{figure}

\begin{figure}[h!]
    \centering
       \centerline{
   \includegraphics[width=18cm]{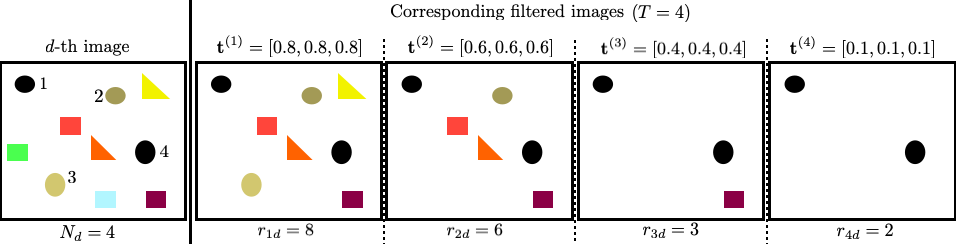}
    }
    \vspace{-0.3cm}
    \caption{{\footnotesize Illustrative example of a generic $d$-th image and $T=4$ corresponding filtered images, with different threshold vectors ${\bf t}^{(k)}$. In this graphical example, we have $N_d=4$ objects of interest (i.e., in our application, microglial cells). The rest of the $6$ objects in the image play the role of irrelevant artifacts. In each filtered image, the total number of objects $r_{kd}$ is given. } }
    \label{FigSUPER_resumen0}
\end{figure}

%%%%%%%%%%%%%%%%%%%%%%%%%%%%%%%%%%%%%%
The high-resolution images obtained in the laboratory contain a huge amount of pixels that are just noise or artifacts. Namely, most of the signal in the input space represents noise in our problem. Moreover, microglial cells are very small and dark objects, with different geometric shapes depending on the specific spinal cord cross-section \citep{Bradesi10,trang_beggs_salter_2011}. For all these reasons, we perform a filtering of the images considering threshold color values.
\newline
Let us consider an RGB image provided by the laboratory, containing possibly the microglial cells\footnote{We recall that the approach described here can be employed for completely different types of application, for instance, with images provided  by a satellite, a telescope, or any other medical machinery.}. Each pixel is represented by 3 color values,
$$
{\bf p}=[p_1,p_2,p_3] \in [0,1]^3,
$$ 
 where each value is in 0 to 1 scale, i.e., $p_j\in [0,1]$. The first value $p_1$ represents the amount of red, $p_2$ represents the amount of green, and $p_3$ represents the amount of blue. Hence, ${\bf p}=[1,1,1]$ represents a white pixel and ${\bf p}=[0,0,0]$ represents a black pixel. We define the vector of threshold color values,
 $$
 {\bf t}=[t_1,t_2, t_3]\in \mathbb{R}^3,
 $$
 in order to build a binary (black and white) {\it filtered image} where only the pixels that satisfy the conditions below
\begin{gather}\label{UmbralesCond}
\left\{
\begin{split}
 p_1 &\leq t_1, \\
 p_2 &\leq t_2, \\
 p_3 &\leq t_3,
 \end{split}
 \right. 
 \end{gather}
 will be considered as black pixels in this binary (black and white) filtered image.
Whereas, all the pixels such that at least one condition is not satisfied in  \eqref{UmbralesCond}, i.e., a $p_i > t_i$, are transformed into white pixels. At each $d$-th image in the database, we apply different threshold vectors ${\bf t}^{(k)}\in [t_1^{(k)},t_2^{(k)}, t_3^{(k)}]$, with $k=1,...,T$. Thus, from each medical colored image in the database from the laboratory we extract $T$ binary filtered images.
The underlying idea is to count the number of black objects (i.e., clusters of black pixels) within each of these filtered images, denoted by the integer variable $r_{kd} \in \mathbb{N}$. In order to count the black objects in this filtered binary image, we can use any kind of clustering algorithm or any alternative procedure designed for counting objects in binary matrices.
\newline
\newline
More specifically, let us assume that we have $D$ images in the database and we consider the $d$-th image to analyze; hence, clearly, $d\in\{1,...,D\}$. Given the $k$-th  threshold vector ${\bf t}^{(k)}\in [t_1^{(k)},t_2^{(k)}, t_3^{(k)}]$, the number of objects within each filtered image is denoted as $r_{kd}\in \mathbb{N}$, i.e., $r_{kd}$ represents the number of objects (clusters of black pixels) in the  $k$-th binary filtered image, obtained filtering $d$-th image i with the threshold vector ${\bf t}^{(k)}\in [t_1^{(k)},t_2^{(k)}, t_3^{(k)}]$. Hence, we have a correspondence between the thresholds and the number of objects in each $d$-th filtered image, i.e., 
$$
{\bf t}^{(k)} \Longrightarrow \{r_{kd}\}_{d=1}^D, \qquad k=1,...,T.
$$
Therefore, to the $d$-th image in the database, we can associate a vector ${\bf r}_d=[r_{1d},r_{2d},...,r_{Td}]$ of the number of objects in each filtered image.
Additionally, for each image in the dataset (provided by the laboratory), the expert provided the expected number of microglial cells $N_d \in\mathbb{N}=\{0,1,2,3,...\}$. Thus, the vector ${\bf r}_d$ is statistically related to $N_d$, i.e., we have the correspondence,
$$
{\bf r}_d=[r_{1d},r_{2d},...,r_{Td}] \Longleftrightarrow N_d.
$$
Therefore, for each medical image in the database $d=1,...,D$, we have $T$ different components of the inputs, $r_{kd} \in \mathbb{N}$, related to the number of microglial cells $N_d$ in the $d$-th image (that plays the role of outputs in a regression problem). 
We now have  a database formed by the pairs inputs/outputs $\{{\bf r}_d,N_d\}_{d=1}^D$, i.e., the counting problem can be tackled as a regression problem. Figure \ref{FigSUPER_resumen_menos1} depicts a graphical sketch of this image analysis.
 Figure \ref{FigSUPER_resumen0} provides a graphical example with $N_d=4$ and $T=4$ different filtered images. Recall that pixels with color values closer to $0$ are darker, whereas  pixels with color values closer to $1$ are clearer.
 Table \ref{TableNotation} summarizes the main notation of the work.
%\newline
%\newline
 
 \begin{table}[ht]
\begin{center}
{\footnotesize
\caption{{\footnotesize Main notation of the work.}}
\label{TableNotation}
\vspace{0.2cm}
\begin{tabular}{|c|c|l|}
\hline
 & &  \\
${\bf t}^{(k)}$ & $k$-th threshold vector & Decided by the user \\ 
 & &  \\
  \hline
   & &  \\
  $N_d$ & number of microglial cells of the $d$-th image in the dataset & Given by the expert\\  
 & &  \\
 \hline
  & &  \\
 $r_{kd}$ &  number of objects in the $k$-th filtered binary image  & Obtained filtering  \\
 & &   the $d$-th image with ${\bf t}^{(k)}$\\
 \hline
  %  \hline
% $\alpha_k$, $s_k$, $v_k$ & coefficients in the observation model in Eq. \eqref{EqModel} & unknown \\
%   $F_k$ & a discrete random variable which takes non-negative values between $v_k$ and $s_k$ & unknown \\
%    \hline
%       \hline 
%  $\#TP$,      & \multirow{2}{*}{true and false positives in an image obtained filtering an image in the dataset} &   \multirow{2}{*}{known by the analysis} \\
%$\#FP$   &   &   \\
 %\hline 
\end{tabular}
}
\end{center}
\end{table}

\noindent
  When the threshold $t_i^{(k)}$ values are small close to zero (i.e., we keep only dark pixels), we have a small number of objects $r_{kd}$ in the filtered image but, since microglial cells are usually formed by dark pixels, most of the $r_{kd}$ objects would be microglial cells, hence we would have a small number of artifacts (false positives). As  the threshold values $t_i^{(k)}$ grow closer and closer to 1, there will be a greater chance of getting a larger portion of the microglial cells in the corresponding filtered image, but also a greater number of artifacts. Furthermore, by still increasing the threshold values, all the microglial cells would be contained in the filtered images. A further increase of the thresholds will only yield an increase in the false positives/artifacts (since all the microglial cells are already contained in the previous filtered images). 
\newline
{
{\bf On the clustering scheme.} For the sake of simplicity, we have used a pre-established Matlab function\footnote{To have more information, see similar functions \url{https://www.mathworks.com/help/images/ref/bwlabeln.html}.} able to count objects  in binary matrices, that are formed by ``islands'' of connected ``1". See figures at \url{https://www.mathworks.com/help/images/label-and-measure-objects-in-a-binary-image.html}, for some examples.
The algorithms are based on the ideas in \cite{Gargantini82,Samet88,Sedgewick98}.  
These algorithms are simple and fast but tend to overestimate the number of objects: for instance, some ``big'' microglial cell can be often counted twice if some part of the cell is  not connected to the rest of the cell by pixels with ``1''. This effect can happen often in high-resolution images. However, if the employed clustering algorithm is always the same for analyzing all the images, the over-estimation effect can be easily corrected by the kernel smoother approach described later (i.e., by the next piece of the proposed counting approach). Hence, several clustering alternatives can be applied (e.g., k-means etc.), but the most important point to remark is that to use the same clustering scheme for analyzing all the images in the dataset is essential. Clearly, 
more precision in the clustering step is welcome and can help the overall proposed strategy.
However, any possible bias introduced by the chosen clustering approach will be corrected (as the number of images in the database grows) by the regression/prediction step included in the proposed counting algorithm (that, in this work, is given by a kernel smoother).

}
  
%%%%%%%%%%%%%%%%%%%%%%%%%%%%%%%%%%%%
\section{A kernel smoother approach for counting microglial cells }\label{SectionCore}
%%%%%%%%%%%%%%%%%%%%%%%%%%%%%%%%%%%%

%As described in the previous section, for each medical image in the database $d=1,...,D$, we have $T$ different components of the inputs, $r_{kd} \in \mathbb{N}$, related to the number of microglial cells $N_d$ in the $d$-th image (that plays the role of outputs in a regression problem). Thus, we can define a vector ${\bf r}_d=[r_{1d},r_{2d},...,r_{Td}]$ which is related to $N_d$, i.e.,
%$$
%{\bf r}_d=[r_{1d},r_{2d},...,r_{Td}] \Longleftrightarrow N_d.
%$$
Let us assume that $D\geq T$.  We also consider a new image that we can  study, obtaining ${\bf r}_{D+1}$, but we do not have the number of microglial cells $N_{D+1}$ given by the expert; thus, we desire to get an estimator $\widehat{N}_{D+1}$. Given the different images in the database and the new image where $N_{D+1}$ is unknown, the database (training samples) is formed by the pairs inputs/outputs $\{{\bf r}_d,N_d\}_{d=1}^D$  and the test input, where we need a prediction, is ${\bf r}_{D+1}$ obtained by analyzing the new image, i.e., we have
\begin{eqnarray*}
        {\bf r}_1&=& [ r_{11},r_{21}, \cdots, r_{T1}]  \Longleftrightarrow N_1, \quad \mbox{ given by the expert},\\
          {\bf r}_2&=& [ r_{12}, r_{22},  \cdots, r_{T2}] \Longleftrightarrow N_2, \quad \mbox{ given by the expert}, \\
         & \vdots& \\
              {\bf r}_D&=& [r_{1D}, r_{2D},  \cdots, r_{Td} ]  \Longleftrightarrow N_D, \quad \mbox{ given by the expert},  \\
       &\mbox{and}& \\        
               {\bf r}_{D+1}&=& [r_{1(D+1)}, r_{2(D+1)},  \cdots, r_{T(D+1)} ]              
\Longleftrightarrow N_{D+1}=?
%          \vdots & \vdots & \vdots  & \vdots  \\
%      r_{1D} & r_{2D} &  \cdots & r_{kd} \\  
%       r_{1(D+1)}^* & r_{2(D+1)}^* &  \cdots & r_{T(D+1)}^* \\  
\end{eqnarray*}
 Then, the goal is to obtain $\widehat{N}_{D+1}$ as a prediction of the number of microglial cells $N_{D+1}$ in the new image that has not been analyzed by the expert. 
Recall that  each component $r_{k(D+1)}$ represents the number of objects in the $k$-th filtered image obtained by filtering the new test image using the threshold vector ${\bf t}^{(k)}$. 

%%%%%%%%%%%%%%%%%%%
\subsection{The kernel counter (KC)} 
%%%%%%%%%%%%%%%%%%%
The proposed algorithm, called {\it kernel  counter (KC)}, is composed of the following steps: (a) standardization, (b) weighting, and (c) prediction,
which are detailed below.
\newline
\newline
{\bf Standardization.} We can standardize each of the $(D+1)$ values for each $t$, as 
\begin{align}
{\bar r}_{kd}=\frac{ r_{kd}-\widehat{\mu}_k}{\widehat{\sigma}_k}, \qquad \widehat{\mu}_k=\frac{1}{D+1}\sum_{d=1}^{D+1} r_{kd},  \qquad  \widehat{\sigma}_k=\sqrt{\frac{1}{D}\sum_{d=1}^{D+1} (r_{kd}-\widehat{\mu}_k)^2},
\end{align}
for all $d=1,...,D+1$.
\newline
\newline
{\bf Weighting.} Denoting as ${\bf {\bar r}}_d=[{\bar r}_{1d}, {\bar r}_{2d},  \cdots, {\bar r}_{Td}] $ the vectors with the standardized values, we can compute the $D$ distances with respect to ${\bar r}_{t(D+1)}$,
\begin{align}
L_d&=||{\bf {\bar r}}_d-{\bf  {\bar r}}_{D+1}||^2=\sum_{k=1}^{T}({\bar r}_{kd}-{\bar r}_{k(D+1)})^2,  \qquad d=1,...,D, 
\end{align}
and the $D$ different weights with the corresponding normalized weights,
\begin{align}\label{predWEIGHTs}
w_d=\exp\left(-\frac{1}{\eta} L_d\right)&=\exp\left(-\frac{1}{\eta}||{\bf {\bar r}}_d-{\bf  {\bar r}}_{D+1}||^2\right), \nonumber \\
&=\exp\left(-\frac{1}{\eta}\sum_{k=1}^{T}({\bar r}_{kd}-{\bar r}_{k(D+1)})^2\right),
\qquad \bar{w}_d=\frac{w_d}{\sum_{i=1}^D w_i}, \qquad d=1,...,D, 
\end{align}
where $\eta>0$ is chosen by the user, or learnt by leave-one-out cross-validation (LOO-CV). Note that, in this weighting step, we convert distances into weights.
 \newline
 \newline
{\bf  Prediction.} The kernel smoother estimator is then given by
 \begin{align}\label{PRED_form}
\widehat{N}_{D+1}= \sum_{d=1}^D \bar{w}_d N_d.
\end{align}
If we desire to get an integer estimation, we can round it, i.e., $\lfloor \widehat{N}_{D+1}\rceil$.
 Note that, even if the formula above is linear, the estimator performs a  non-linear regression  with respect to the input vectors ${\bf {\bar r}}$ \cite{MARTINOfusion}.
An estimation of the variance $\widehat{\sigma}^2_{D+1}$ associated to $\widehat{N}_{D+1}$ is given in Eq. \eqref{VAR_PRED_form} below (see also \cite{LeonPaper}).
%%%%%%%%%%%%%%%%%%%%%%
%\subsection{Properties of the KC estimator}
%%%%%%%%%%%%%%%%%%%%%%
\newline
\newline
The KC estimator has some interesting properties that we discuss below.
\newline
\newline
{\bf Property 1.} Note that $\widehat{N}_{D+1} \geq 0$, which is a desired property for a counting algorithm. This is because $\widehat{N}_{D+1}$ is obtained as a linear combination of non-negative quantities, i.e.,  $N_d  \geq 0$, and all the weights are also non-negative, i.e.,  $\bar{w}_d  \geq 0$.
\newline
\newline
{\bf Property 2.}  We have that $\min\limits_d N_d \leq  \widehat{N}_{D+1} \leq \max\limits_d N_d$. Therefore, the increase in the database is also beneficial for increasing the prediction ability and flexibility of the algorithm.
\newline
\newline
{\bf Remark.} In this version of the algorithm, we have a unique hyperparameter to learn that is the non-negative scalar $\eta$. It can be learnt by leave-one-out cross-validation (LOO-CV). For this reason, the proposed KC is easy and fast to train. However, other possible KC versions (with more hyper-parameters) are discussed in the next sections below.
\newline
\newline
{\bf Remark.} Even if we have only one hyperparameter to learn, the method is a {\it non-parametric} regressor, i.e., the  complexity of the solution in Eq. \eqref{PRED_form} grows with $D$ (that is, the number of images in the database). Moreover, the solution is {\it non-linear}  with respect to the inputs ${\bf \bar{r}}$. Hence, the proposed method is able to express the complexity of rich datasets. This kernel procedure allows also the estimation of the variance $\widehat{\sigma}^2_{D+1}$ given  in Eq. \eqref{VAR_PRED_form}.  Furthermore, the extensions with multi-expert's opinions is straightforward as shown in the next section.

%%%%%%%%%%%%%%%%%%%%%%%%%%%%
\subsection{Smoothing of the expert's opinions and variance of the prediction}
%%%%%%%%%%%%%%%%%%%%%%%%%%%%
The KC algorithm can also be used to propose a {\it correction} of the expert's opinions, given all the information in the dataset.
 Generalizing the previous formulas, changing the reference vector,
we have
\begin{align}
L_{dj}&=||{\bf {\bar r}}_d-{\bf  {\bar r}}_{j}||^2=\sum_{k=1}^{T}({\bar r}_{kd}-{\bar r}_{kj})^2,  \qquad d=1,...,D, \quad j=1,...,D.
\end{align}
Note that $L_{jj}=0$ for all $j$.
Fixing  now $j$, we can again define $D$ different weights and the corresponding normalized weights,
\begin{align}
\rho_{dj}&=\exp\left(-\frac{1}{\eta} L_{dj}\right),
\qquad \bar{\rho}_{dj}=\frac{\rho_{dj}}{\sum_{i=1}^D \rho_{ij}}, \qquad d=1,...,D.
\end{align}
Note that $\rho_{kj}=\rho_{jk}$ since $ L_{kj}=L_{jk}$.
Note that $w_d=\rho_{d(D+1)}$ for all $d$, i.e., if we use ${\bf \bar{r}}_{D+1}$ as reference vector, we recover the previous weights in Eq. \eqref{predWEIGHTs}, as expected.
The smoothing of the expert's opinion for the $j$-th image is 
 \begin{align}\label{smoothEQ}
\widehat{N}_{d}= \sum_{k=1}^D \bar{\rho}_{kd} N_k, \qquad d=1,...,D.
\end{align}
%See Figure  \ref{FigSUPER_resumen}, for an example of construction of the vectors ${\bf r}_d$, {\bf c} and the matrix ${\bf R}$.
{\bf  Variance of the prediction.} Considering the smoothing values  $\widehat{N}_{d}$ computed above,  we can also estimate the variance in the prediction $\widehat{N}_{D+1}$ in \eqref{PRED_form} as suggested in \cite{LeonPaper}, i.e.,  
 \begin{align}\label{VAR_PRED_form}
\widehat{\sigma}^2_{D+1}= \sum_{d=1}^D \bar{w}_d \left(N_d- \widehat{N}_{d}\right)^2.
\end{align}
Note that this estimation of the variance is obtained directly by applying a formula without any bootstrap or similar procedures. 
In the same fashion, the variances of the smoothing values $\widehat{N}_{d}$ can be approximated by
 \begin{align}\label{VAR_PRED_form2}
  \widehat{\sigma}^2_{d}= \sum_{k=1}^D \bar{\rho}_{kd} \left(N_k- \widehat{N}_{k}\right)^2.
  \end{align}
  Estimation of correlations or higher moments could also be provided as suggested in \cite{LeonPaper}.

%%%%%%%%%%%%%%%%%%%%%%%%%%%%%%%%%
\section{Learning $\eta$ and extension with more hyper-parameters}
\label{EtaSect}
%%%%%%%%%%%%%%%%%%%%%%%%%%%%%%%%%

The described KC  has a unique hyperparameter $\eta$.
Note that $\eta$ controls the underfitting/overfitting trade-off. Indeed, for instance, for big values of $\eta$ we tend to the underfitting. In the limit case of  $\eta \rightarrow \infty$,  we have that 
$$
 \bar{w}_d =\frac{1}{D},  \quad  \widehat{N}_{D+1}= \frac{1}{D}\sum_{d=1}^D  N_d,
$$
so that the prediction is just the arithmetic mean of the number of microglial cells, $N_d$, in the different images. As $\eta$ decreases,  we tend to overfit. In the limit case of  $\eta \rightarrow 0$,  we have that $\widehat{N}_{D+1}$ is given by the number of microglial cells  $N_{d^*}$ corresponding to the nearest neighbor solution, i.e., $\widehat{N}_{D+1}=N_{d^*}$, where
$$
d^*=\arg \min_d ||{\bf {\bar r}}_d-{\bf  {\bar r}}_{D+1}||, \qquad d=1,...,D.
$$
In this scenario, only the value $N_{d^*}$ is taken into account, i.e., $ \bar{w}_{d^*}=1$, whereas the rest of the weights are zero, $\bar{w}_{d}=0$ for any $d\neq d^*$. Hence, the KC contains the nearest neighbor algorithm as a special case (see \cite{MARTINOfusion} for more details).
\newline
\newline
An optimal value  $\eta^*$ can be obtained by LOO-CV, trying to minimize an error loss between the predicted and true numbers of cells, or maximizing the coefficient of determination of the linear regression  between the predicted and true numbers of cells (any other metric within LOO-CV can be employed). Note that, in any case, as the size $D$ of the dataset grows, the optimal value of $\eta^*$ decreases, i.e., as $DÊ\rightarrow \infty$ then $\eta^*\rightarrow 0$.
As a good starting point for the LOO-CV procedure (to understand the order of magnitude of $\eta$), or as a reasonable proxy of $\eta^*$, one can use the following rule of thumb:
\begin{align}
\widehat{\eta}= \frac{2}{DT}\sum_{d=1}^D \sum_{k=1}^{T}({\bar r}_{kd}-{\bar r}_{k(D+1)})^2,
\end{align}
based on the empirical estimator of a variance. This is based on the fact that  we are employing Gaussian kernels, and $\frac{1}{2}\widehat{\eta}$ plays the role of a variance parameter.  The value $\widehat{\eta}$ tends to the under-fitting for big values of $D$ (i.e., $\eta^* < \widehat{\eta}$), whereas  is already a reasonable value for small $D$.
\newline
As already remarked above, even with just one hyperparameter to learn, the method is a {\it non-parametric} and {\it non-linear} regressor, with respect to the input vectors ${\bf \bar{r}}$. The complexity of the final estimator in Eq. \eqref{PRED_form} grows with $D$.
However, even more flexible versions of KC, with more hyperparameters, can be easily designed. For instance, we can have one  hyperparameter $\eta_d$ for each $d$-th image and, as a consequence, for each weight $w_d$ as shown below,
\begin{align}
w_d=\exp\left(-\frac{1}{\eta_d} ||{\bf {\bar r}}_d-{\bf  {\bar r}}_{D+1}||^2\right), \qquad d=1,...,D.
\end{align}
Therefore, in this scenario, we have $D$ hyperparameters $\eta_d$, for  $d=1,...,D$, i.e., also the number of hyperparameters grows with the number of data $D$ in the dataset. The rules of thumb in this scenario would be $\widehat{\eta}_d= \frac{2}{T} \sum_{k=1}^{T}({\bar r}_{kd}-{\bar r}_{k(D+1)})^2$, for any $d$.
\newline
\newline
{ 
{\bf Possible alternative regression schemes.} The kernel smoother approach described here meets all the requirements outlined for Part P2 in Section \ref{KeySect}.  Additional discussion on its ability to incorporate expert uncertainty is provided in Section \ref{RobustenessSect}. Recall that 
the complete proposed framework consists of two main components: feature extraction (P1) and a regression step (P2).
Practitioners may replace the regression module in P2 with a deep learning approach (or any other alternatives), if it is considered more appropriate,  and if  all the requirements of P2 given in Section \ref{KeySect} are fulfilled.   
 }

%%%%%%%%%%%%%%%%%%%%%%%%%%%%%%%%%%%%
\section{Increasing the robustness of the KC}\label{RobustenessSect}
%%%%%%%%%%%%%%%%%%%%%%%%%%%%%%%%%%%%

In the data-collecting process, the quality of the obtained image can vary according to the employed  imaging systems and the human activities that can yield distortion or information loss. For instance, an image can be obtained in different lighting conditions. Moreover, the counting procedure by the expert is also a noisy process. Recall also that the database is quite small since the analysis of the expert's opinion is costly. In this section, we discuss strategies to handle these issues and  improve the robustness of the KC.

{
\subsection{Adaptive thresholds} 
}

The validity of the proposed algorithm is ensured for any possible choice of ${\bf t}^{(k)}$, with the unique requirement that the vectors ${\bf t}^{(k)}$, with $k=1,...,T$, must be different from each other. However, the performance of the algorithm depends on the choice of threshold vectors. See Appendix \ref{SectionChoice} for more details.
In order to reduce the sensitivity of the image conditions (such as filter type, lighting conditions, etc.), we suggest to covert the threshold values into areas below the approximated densities of each color, and then into the quantile values of each new image. 
Namely, choosing the threshold vector ${\bf t}=[t_1,t_2, t_3]$, we can convert it into a probability vector ${\bf a}$,
 \begin{align}
  {\bf a}=[a_1,a_2, a_3], \quad\mbox{ where } \quad a_k\approx \mbox{Prob(a pixel  has the $k$-color $\leq t_i$)},
 \end{align}
obtained by studying all the histograms of the colors of all the images in the database. Then, given a new test image, analyzing its histograms of color we can compute the approximated quantile values, 
 \begin{align}
 q_k \approx \mbox{quantile of order $a_k/100$}, \quad \mbox{for \quad $i=1,2,3$}.
  \end{align}
Finally, the corresponding threshold vector for analyzing the new image will be 
$$
{\bf t}_{\texttt{new}}=[q_1,q_2, q_3],
$$
which is the new vector of the thresholds {\it adapted} to the new image.
This procedure improves the robustness of the KC by analyzing darker or lighter images (due to experimental changes in the laboratory) included in the same dataset. %reducing the dependence of the chose
Hence, the adaptive threshold strategy described above has another advantage: it allows us to include the same image {\it but} with a different degree of clarity for increasing the size dataset. This is a straightforward data augmentation procedure that can be used that also increase the robustness of the proposed algorithm.

{
\subsection{Including the expert's uncertainty} 

One characteristic of the tackled problem  is the presence of noisy labels, resulting from a laborious visual inspection performed by a human expert. This introduces structural uncertainty, which we aim to address without increasing the complexity of either the dataset creation process or the training of the algorithm. Below, we describe some strategy easily to apply within the KC.
}
\newline
\newline
{{\bf Strategy-1.}} The first simple possibility for including the expert's uncertainty (within the proposed KC algorithm) is to consider a ``soft'' labeling approach. Namely, in the $d$-th image of the dataset, let us consider that we have a  certain number of objects, $O$, which can be considered possibly a microglial cell by  the expert. For each one of these objects, the expert can associate  a value $p_{o}$ between $[0,1]$ (as a probability) where values close to zero, i.e., $p_{o} \approx 0$, represent very high uncertainty that the object is a microglial cell, whereas $p_{o}=1$ means a complete guarantee that this $o$-th object is a microglial cell. 
 In this framework, finally, we have
$$
N_d=\sum_{o=1}^O p_{o}\in \mathbb{R}^+, \qquad  p_{o}\in [0,1],
$$ 
which is a positive real number, $N_d\in \mathbb{R}^+$, representing the number of microglial cells in the $d$-th image of the dataset, including the expert's uncertainty. 
The rest of the algorithm would remain exactly the same. 
\newline
\newline
{ {\bf Strategy-2.}} Another possibility to encode the expert's uncertainty is to include some additional weights $\beta_d \in [0,1]$ directly in the final KC estimator
\begin{align}\label{PRED_form_2}
\widehat{N}_{D+1}= \sum_{d=1}^D \bar{w}_d \ \beta_d  N_d,
\end{align}
 where if $\beta_d=0$ represents maximum uncertainty, so that $N_d$ is not considered in the linear combination, whereas   if $\beta_d=1$ represents zero uncertainty to the value $N_d$ computed  by the expert. 
  { In any case, the proposed KC always returns a non-negative prediction/estimation.} 
 \newline
 \newline
 { 
 {\bf Strategy-3.} Another way to include the expert's uncertainty is to ask for providing lower and upper bounds for each image, i.e., $N_{d}^{\texttt{Low}}$ and  $N_{d}^{\texttt{Upp}}$, so that, following the expert opinion, we have $N_{d}^{\texttt{Low}}\leq N_d \leq N_{d}^{\texttt{Upp}}$. We can directly apply the kernel smoother approach to both, the lower values and to the upper values, making predictions also  to this interval of values (and  providing uncertainty estimation on each one of these values). Another procedure to handle this situation is to consider it a two-output scenario,
 $$
     {\bf r}_d  \Longleftrightarrow {\bf N}_{d}=[N_{d}^{\texttt{Low}},N_{d}^{\texttt{Upp}}],
 $$ 
 As we will explain below, this two-output scenario can be directly handled by the KC formulas above, giving in this case a unique value as estimator $\widehat{N}_{D+1}$.
 Note that this of including the expert's uncertainty can facilitate the database creation,  making it faster and easier to label the images by the expert. }
\newline
\newline
{\bf Several experts.} Let us consider that different experts (more than one specialist) provide their count of the microglial cells of the same image. Namely, for the $d$-th image a $j$-th expert suggests that the number of microglial cells is $N_{dj}$. Moreover, let us consider that the total number of experts is $E$. Therefore, after the filtering, we have the correspondence:
$$
     {\bf r}_d  \Longleftrightarrow {\bf N}_{d}=[N_{d1},N_{d2},...,N_{dE}],
 $$ 
 i.e., we have a vector of outputs in the regression problem. This multi-output scenario can be directly handled by the KC just considering the different pairs $\{{\bf r}_d, N_{dj}\}$ for $j=1,...,E$ and $d=1,...,D$ as different data points \cite{MARTINOfusion}. The total number of data will be then $ED$, and the presented formulas can be applied directly.
 
{
\subsection{ Data augmentation}
\label{dataAug}
When we talk about the expert's uncertainty, we are referring to the uncertainty in the output values.
We can also address the uncertainty in the input vectors (due to the varying characteristics  of imaging acquisition systems used across different laboratories) by applying a data augmentation strategy.
Indeed, the proposed algorithm can also be trained using artificially generated data, obtained by applying controlled variations to the labeled images. 
\newline
For instance, the user may artificially darken or lighten existing labeled images (by simply applying a constant translation to the pixel intensity values), and include them in the training dataset.
Other simple approaches based on the output uncertainties are also possible. Recall that we have $D$ (true) labeled images in the database.  
Another idea is to re-include all the possible input vector ${\bf r}_d$ in the dataset as a new $j$-th element ($j>D$), where we have $\frac{1}{2}(N_{d}^{\texttt{Low}}+N_{d}^{\texttt{Upp}}) \neq N_d$,
but with a different output defined as
\begin{align}
    N_{j}=\frac{1}{2}(N_{d}^{\texttt{Low}}+N_{d}^{\texttt{Upp}}),
\end{align}
and ${\bf r}_j={\bf r}_d$, with $j>D$. The data augmentation strategy increases the robustness of the algorithm by effectively expanding the diversity and size of the training dataset. Namely, it facilitates database creation by reducing the manual effort required from the human operator.
}

%%%%%%%%%%%%%%%%%%
{
\subsection{All the elements and their sensitivity}
%%%%%%%%%%%%%%%%%%

The elements of the overall designed method are:
\begin{enumerate}
\item {\bf The threshold vectors   ${\bf t}^{(k)}$.} The most important parameters are the threshold values for filtering the images. First of all, it is important to remark that the proposed KC algorithm provides consistent results for any choice of ${\bf t}^{(k)}$, and $T\rightarrow \infty$, as shown in Figure \ref{FigMSEvTb}. Moreover, the MSE can be very small even with a finite $T$; again, see Figure \ref{FigMSEvTb}. However, the performance is highly influenced by the choice of the vectors ${\bf t}^{(k)}$. In fact, even with a small $T$, selecting appropriate ${\bf t}^{(k)}$ vectors can lead to excellent results. The careful selection  of  ${\bf t}^{(k)}$ is discussed in Appendix \ref{SectionChoice}. The threshold values should be chosen to maximize the ratio of true microglial cells to false positives in the filtered images.

\item {\bf The total number of threshold vectors, $T$.}  
Clearly, increasing $T$ leads to equal or improved results. Generally, we have an improvement (smaller error), but  the error may remain stationary and not decrease further if the additional inputs consist primarily of noise and do not contain ``true signal'' (i.e., information). This is the reason why the most important parameters are the threshold vectors ${\bf t}^{(k)}$, with $k=1,...,T$.

\item  {\bf The hyperparameter $\eta$.} Based on our experience, we find that tuning this parameter using LOO-CV is relatively straightforward. Generally, as the dataset grows, there tends to be a range of $\eta$ values that yield good and similar results. In contrast, with smaller datasets, the cost function produced by LOO-CV often exhibits a more peaked behavior. Nevertheless, since we are dealing with a one-dimensional optimization problem, a finer search grid or any other suitable method can be employed to identify the optimal $\eta$. Moreover, the ``rule of thumbs'', given in Section \ref{EtaSect}, usually provides good values of $\eta$, at least for an initialization. Some additional considerations are given in Section \ref{ResSubsect}.

\item {\bf Type of kernel function.} Other different kernel functions could be applied but, in our opinion, the impact on the results is minimal with respect to the variation of   other elements.
\end{enumerate}
}

%}

%%%%%%%%%%%%%%%%%%%%%%%%%%%%%%%%%
%\section{Application to microglial cells}
%%%%%%%%%%%%%%%%%%%%%%%%%%%%%%%%%

{
%%%%%%%%%%%%%%%%%%%%%%%%%%%%%%
\section{Benefits with respect to other approaches}\label{OtherApproaches}
%%%%%%%%%%%%%%%%%%%%%%%%%%%%%%%%%%%%

The overall introduced scheme, including both parts P1 and P2 as described in Section \ref{KeySect}, offers several advantages for the specific task of microglial cell counting,  when compared to alternative techniques proposed in the literature (e.g., deep learning methods). The main benefits are summarized below:

\begin{itemize}
\item[(a)] The dataset may include heterogeneous images acquired under diverse conditions and circumstances, and may also differ in size. Moreover, only portions of images can be included, which further simplifies the expert's labeling task, as different regions of the same image can be incorporated at different stages.
\item[(b)] Also related to the previous point, the manual creation of the dataset is significantly simplified: the proposed scheme does not require pixel-level annotations, but only the total number of cells within each analyzed image region.
This is because our approach is designed to perform counting, not object detection.

\item[(c)] The proposed scheme is also able to manage and include uncertainty of the labels/outputs,  i.e., to deal with with noisy labels/outputs due to potential errors or 'doubts' introduced by the expert during the counting process. This further simplifies the dataset creation, allowing the expert to perform the counting more quickly without the concern of being perfectly precise.

\item[(d)] The KC algorithm always produces {\it non-negative predictions}, effectively functioning as a counter (even if trained in a small database).
\item[(e)] The training of the KC is easy and fast with both small or great datasets. At the same time, it offers strong expressive power. In fact, the proposed algorithm in P2 is a non-parametric regression method that is inherently flexible and, if needed, capable of fully overfitting the data.
\item[(f)] The KC algorithm (part P2) is also able to provide the {\it uncertainty estimation} of the given predictions. 
Since KC also smooths the expert's opinion (i.e., reduces noise in the outputs $N_d$), a high uncertainty in the smoothed value may indicate that the image requires revision by the expert.
\item[(g)] The first part (P1), i.e., the tailored feature extraction, serves to eliminate irrelevant information from the high-dimensional images, generating new inputs with a significantly higher signal-to-noise ratio. These transformed inputs retain the essential information in a much lower-dimensional space (allowing to achieve good performance).
\end{itemize}
As highlighted in points (a)-(b)-(c) above, the proposed scheme simplifies the database creation by reducing the manual effort required from human operators, 
making it particularly suitable for laboratories with limited funding and resources that wish to build their own datasets and carry out independent studies. Conversely, neural network and deep learning-based approaches usually depend on pixel-level annotations of the images, which are labor-intensive to produce and crucial for precise localization of the structures of interest (i.e., the microglial cells). 
}

%%%%%%%%%%%%%%%%%
\section{Numerical experiments}\label{SectionSimu}
%%%%%%%%%%%%%%%%%

In this section, we test the KC algorithm in two synthetic experiments, where we can check and verify the convergence behaviors of the proposed scheme (since the ground-truth values are known). Then, in the last section, we apply the KC algorithm to the dataset obtained by our  laboratory. Related Matlab code is also provided.

% We will test with a  progressively closer to real-world application
%%%%%%%%%%%%%%%%%%%%%%%%%%%%%%%
\subsection{First synthetic experiment}
%%%%%%%%%%%%%%%%%%%%%%%%%%%%%%%%%%%%%%%%%%%%%
\label{STATmod_2}
%%%%%%%%%%%%%%%%%%%%%%%%%%%%%%%%%%%

\begin{figure}[!h]
\begin{center}
\centerline{
\subfigure[\label{FigMSEvTa}]{\includegraphics[scale=0.5]{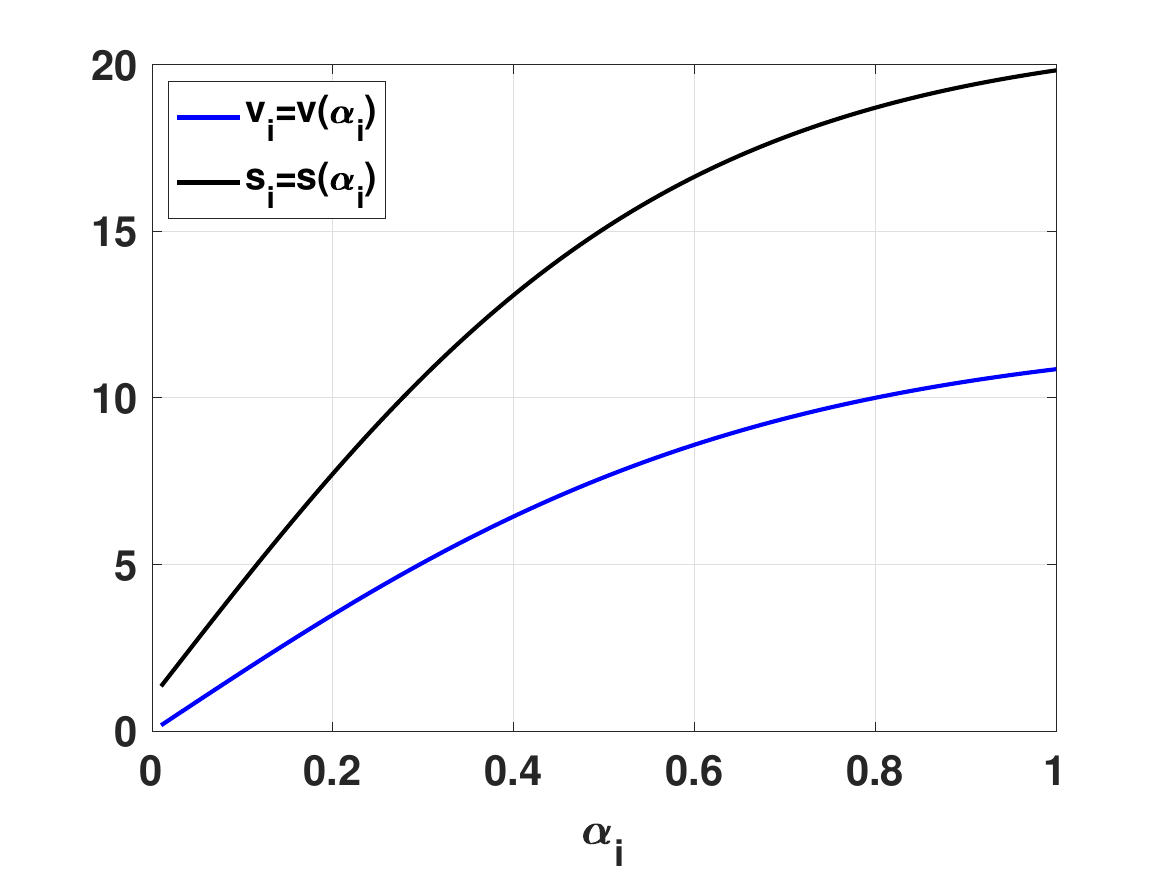}}
\subfigure[\label{FigMSEvTb}]{\includegraphics[scale=0.5]{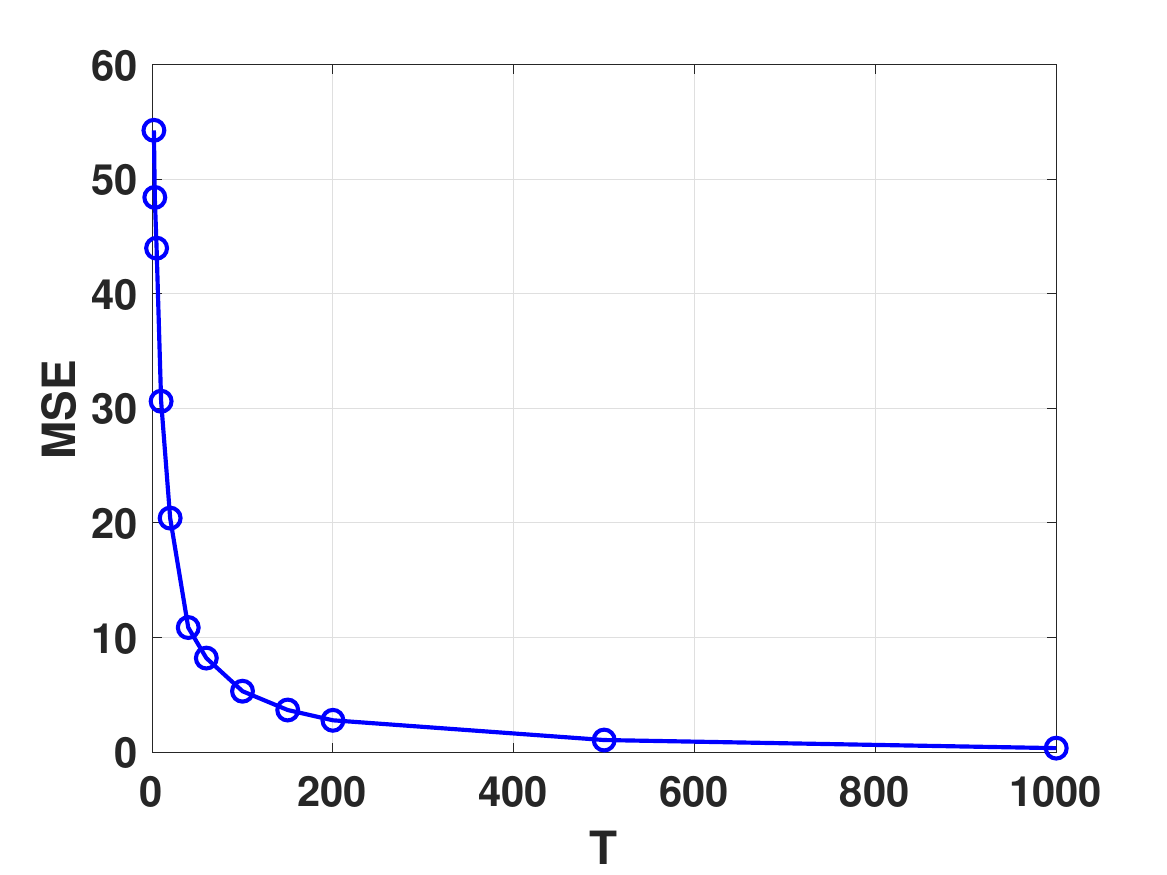}}
}
\caption{{\bf (a)} The two functions $v_i=v(\alpha_i)$ and $s_i=s(\alpha_i)$ in Eqs.Ê\eqref{Vieq} and \eqref{Sieq}. Recall that $\min F_i=v_i$ and $\max F_i=s_i$. {(\bf b)} The mean square error (MSE) in Eq. \eqref{MSEEq} as function of the number of threshold vectors used, i.e., $T$.   }
\label{FigMSEvT}
\end{center}
\end{figure}

In Appendix \ref{STATmod}, we provide a statistical model that can be employed for synthetic experimental analysis and study the theoretical behavior of the proposed KC algorithm. Thus, we consider the statistical model presented in Appendix \ref{STATmod}, i.e.,  
  \begin{gather}
    \begin{split}
  \label{ModelAgain}
r_{id}&=\lfloor  \alpha_i N_d+F_i \rceil, \quad \alpha_i\in (0,1], \quad i=1,...,T, \mbox{ }\mbox{ }d=1,...,D, \\
F_i &\in\{v_i,v_i+1,...,s_i\} \quad s_i,v_i\in \mathbb{N}^+, \quad s_i>v_i.
 \end{split}
 \end{gather}
 where $\lfloor b \rceil$ returns the nearest integer to $b$; the coefficient $\alpha_i\in (0,1]$  depends on the vectors ${\bf t}^{(i)}$, i.e., $\alpha_i=\alpha({\bf t}^{(i)})$, and $F_i$ is a discrete random variable which takes non-negative integer values, contained in  $[v_i, s_i]$.
%$F\in\mathbb{N}=\{0,1,2,3,...\}$
 Eq. \eqref{ModelAgain} shows the relationship between each $r_{id}$ and $N_d$. Indeed,  the value $100 \alpha_i \%$ is the percentage of microglial cells $N_d$ in the $i$-th filtered image (i.e., percentage of true positives), whereas $F_i$ is the number of  false positives (artifacts).
 We consider that in each $d$-th image, we can have a number of microglial cells between 0 and $200$, i.e., $N_d\in [0,200]$. More precisely, we consider a uniform discrete pmf,
  \begin{align}\label{ModelAgain2}
N_d\in \mathcal{U}_{\texttt{discrete}}([0,200]).
 \end{align}
In this section, we assume that the expert is able to detect {\it perfectly} all the number of microglial cells $N_d$ in the $d$-th image, observing the values $N_1,...,N_D$.
\newline
Recall that the value $\alpha_i\cdot 100$ represents the percentages of microglial cells in the $i$-th filtered image (see App. \ref{STATmod}), obtained by filtering the $d$ images with the i-th threshold vector ${\bf t}^{(i)}$.
We consider $T$ threshold vectors ${\bf t}^{(i)}$ such as
$$
\alpha_i =\frac{i-1}{T-1}, \qquad i=1,...,T,
$$
i.e., we assume that ${\bf t}^{(1)}$ is chosen such that $\alpha_1=0$,  ${\bf t}^{(2)}$ is chosen such that $\alpha_2=\frac{2}{T-1}$, ${\bf t}^{(3)}$ is chosen  such that $\alpha_3=\frac{3}{T-1}$ and so on, until ${\bf t}^{(T)}$ that is chosen such that $\alpha_T=\frac{T-1}{T-1}=1$. Hence, we have always $\alpha_1=0$ and $\alpha_T=1$.  
 Moreover, we assume a uniform discrete distribution for the random variable $F_i$,
$$
F_i \sim \mathcal{U}_{\texttt{discrete}}([v_i,s_i]) \in \mathbb{N},
$$
where we can write the minimum and maximum values as function of $\alpha_i$, for instance,
\begin{align}
v_i&=\left\lfloor 24\left(\frac{1}{1+\exp(-3 \alpha_i)}\right)-12\right\rceil, \label{Vieq}  \\
s_i&=\left\lfloor 1+40\left(\frac{1}{1+\exp(-3.5 \alpha_i)}\right)-20 \right\rceil.  \label{Sieq}
\end{align}
Figure \ref{FigMSEvTa} depicts these curves.
 Note that this is equivalent to be function of ${\bf t}^{(i)}$. We generate $N_d$ and $r_{id}$ by the model above, assuming $D=10^4$ images in the dataset, $i=1,...,D$. In this synthetic experiment, we are considering a high value of $D$ since we are studying the convergence behavior of the proposed algorithm. We also consider different values of 
$$
T\in\{2,3,5,10,20,40,60,100,150,200,500,1000\},
$$
i.e., different numbers of  threshold vectors ${\bf t}^{(i)}$ (and consequently generating different filtered images).  
Since in this artificial experiment, we know the ground-truth values $N_1,...,N_D$, and we can compute the mean square error (MSE) in smoothing, i.e.,
\begin{align}\label{MSEEq}
\text{MSE}(T)=\frac{1}{D}\sum_{d=1}^D(\widehat{N}_d-N_d)^2.
\end{align}
We have averaged all the results over $10^5$ independent runs. We employ  LOO-CV for learning $\eta$ at each run.  
   Figure \ref{FigMSEvTb} depicts the results.  We can observe that the MSE is decreasing quickly as $T$ grows. Note that when $T=200$, the MSE is already virtually zero.  %In a new test image, where the number microglial cells and the number of false alarms follow the model given above in Eqs. \eqref{ModelAgain}, \eqref{ModelAgain2},  \eqref{Vieq} and \eqref{Sieq}, the MSE in prediction will be very close to the values obtained in Figure \ref{FigMSEvTb}. 

%%%    2 --- 6.0794 --- 54.2638
 %%%    3 --- 5.6561 --- 48.4068
 %%%    5 --- 5.3947 --- 43.9775 
 %%%   10 --- 4.4514 --- 30.6195
 %%%   20 --- 3.6069 --- 20.4195
 %%%   40 --- 2.6149 --- 10.8707
 %%%   60 --- 2.2574 ---  8.2037
 %%%  100 --- 1.8085 ---  5.3122
 %%%  150 --- 1.4970 ---  3.6714
 %%%  200 --- 1.2909 ---  2.7733
 %%%  500 --- 0.7553 ---  1.0516
 %%% 1000 --- 0.3323 ---  0.3433

%%%%%%%%%%%%%%%%%%%%%%%%%%%%%%%%%%%%%%%%%%%%%%%%%
\subsection{Second synthetic experiment}
\begin{figure}[!h]
\begin{center}
\centerline{
%\subfigure[\label{FigMSEvT_nonLin}]{\includegraphics[scale=0.5]{FIG_MSE_versus_T_nonlin}}
\subfigure[\label{FigMSEvTotro}]{\includegraphics[scale=0.5]{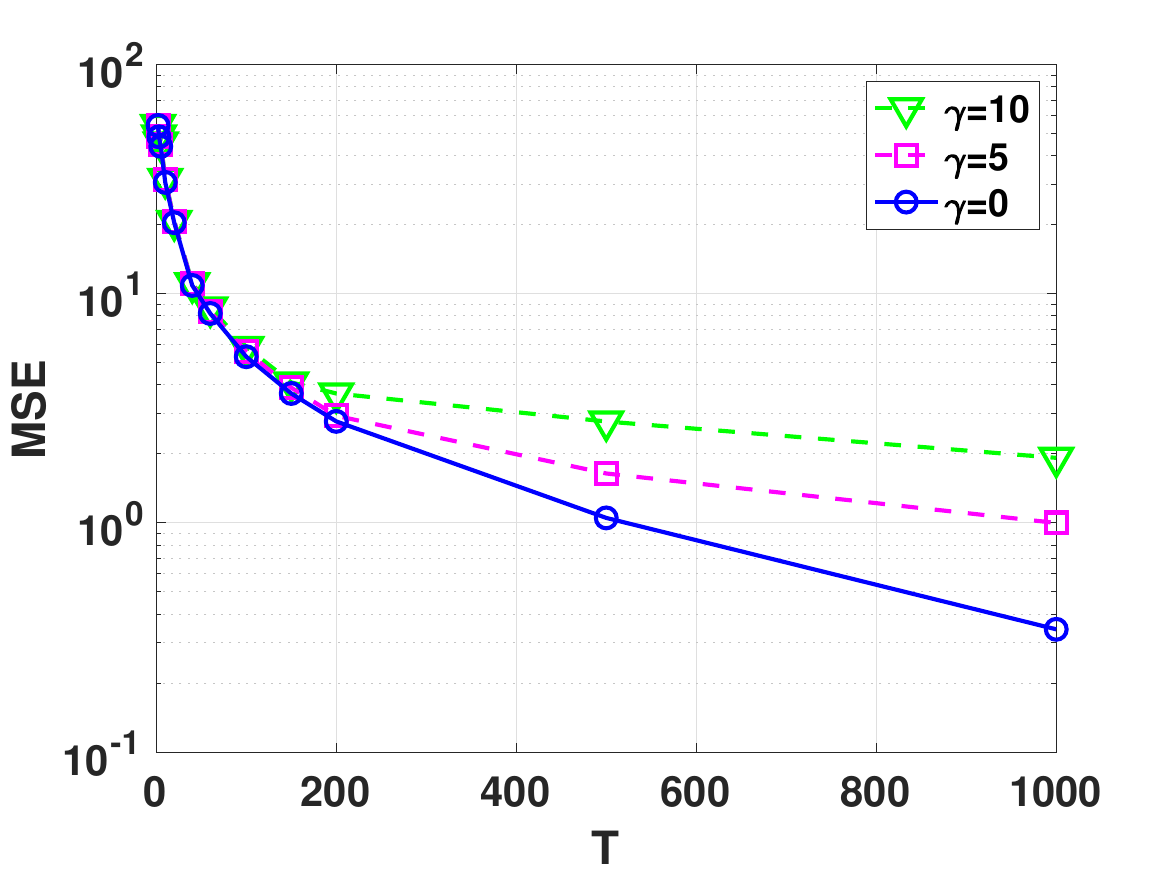}}
\subfigure[\label{FigMSEvTotro2}]{\includegraphics[scale=0.5]{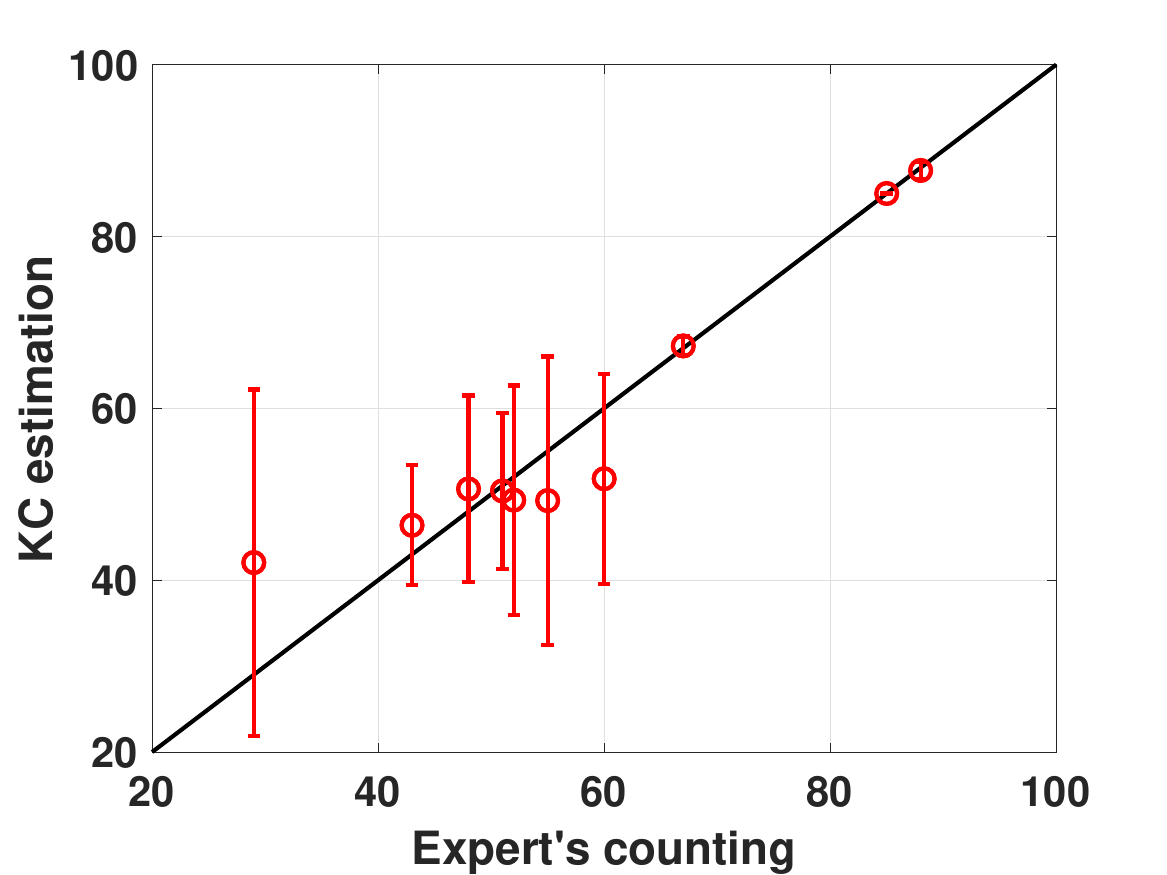}}
}
\caption{
 {\bf (a)} The MSE curve in log-scale as a function of the number $T$ of used threshold vectors, and different values of $\gamma=0,5,10$. Note that the curve in solid line corresponds $\gamma=0$ (i.e., without noisy evaluations of $N_d$) and is the same as in the curve in Figure \ref{FigMSEvTb} but in log-scale. For any value of $\gamma$, the MSE of the algorithm vanishes to zero as $T$ grows, regardless of the power of the noise that affects $N_d$. {\bf (b)}  Results applying the KC to the  real dataset. After LOO-CV, we get $\eta^*\approx 2.5$. The coefficient of determination  between the numbers of cells given by the expert and the predicted numbers is $R^2\approx 0.90$.  The KC is also able to provide $\widehat{\sigma}_d^2$: the error bars show $\pm 2\sqrt{\widehat{\sigma}_d^2}$ corresponding to the $95\%$ of the probability. Note that the error bar always contains the black line, hence the expert's opinion is always contained in the uncertainty interval.    }
\label{FigMSEvTotro_all}
\end{center}
\end{figure}

%%%%%%%%%%%%%%%%%%%%%%%%%%%%%%%%%%%%%%%%%%%%%%%%%
In this section, we consider a more complex version of  the previous  model,
  \begin{gather}
    \begin{split}
  \label{ModelAgain_cazzo}
r_{id}&=\lfloor  \alpha_i N_d+F_i \rceil, \quad \alpha_i\in (0,1], \quad i=1,...,T, \mbox{ }\mbox{ }d=1,...,D, \\
F_i &\in\{v_i,v_i+1,...,s_i\} \quad s_i,v_i\in \mathbb{N}^+, \quad s_i>v_i, \\
 \widetilde{N}_d&=N_d+\epsilon_d, \qquad  \epsilon_d \sim  \mathcal{U}_{\texttt{discrete}}([-\gamma,\gamma]),
 \end{split}
 \end{gather}
 where $\epsilon_d \in [-\gamma,\gamma]$ is a uniform noise variable and $\gamma$ is a constant integer parameter that indirectly determines the power of this noise perturbation. We assume to  observe a noisy measurement $\widetilde{N}_d$, i.e.,  
instead of the true number of microglial cells $N_d$, as in the previous experiment, so that that the data pairs
$$
\{{\bf r}_{d},\widetilde{N}_d \}_{d=1}^D,
$$ 
 where ${\bf r}_{d}=[r_{1d}, ...,r_{Td}]$.
Namely, in this experiment, we are assuming that the expert provides {\it noisy} versions $\widetilde{N}_d $ of the number $N_d$ of microglial cells in the $d$-th image, so that we can test the robustness of the KC algorithm. In this experiment, we have considered $\gamma \in\{0,5,10\}$. Therefore, we use  $\widetilde{N}_d$ in the estimators  in Eqs. \eqref{PRED_form} and \eqref{smoothEQ}. Note that, setting $\gamma=0$, we recover the non-noisy framework in the previous section.  We fix $D=10^4$ and use  LOO-CV for learning $\eta$ at each run. We consider the same equations \eqref{ModelAgain2}, \eqref{Vieq}, and  \eqref{Sieq} for the variables of the model above. 
 \newline
 \newline
 The results are averaged over $10^5$ independent runs.
At each run, we compute the MSE that is still computed considering  the true corresponding values $N_1,...,N_D,$  
exactly as in Eq. \eqref{MSEEq}, i.e., 
\begin{align}\label{MSEEq2}
\text{MSE}(T)=\frac{1}{D}\sum_{d=1}^D(\widehat{N}_d-N_d)^2.
\end{align}
In this way, we can appreciate the deterioration of the performance of the algorithm as $\gamma$ grows, as shown in Figure \ref{FigMSEvTotro}, where the MSE curve is depicted in log-scale to facilitate the visualization. However, the MSE of the algorithm vanishes to zero as $T$ grows, regardless regardless the noise power. This proves the robustness of the proposed KC scheme.

%%%%%%%%%%%%%%%%%%%%%%%%%%%%%%%%%%%%%%%%%%%%%%%%%
\subsection{Counting microglial cells in a real dataset}
%%%%%%%%%%%%%%%%%%%%%%%%%%%%%%%%%%%%%%%%%%%%%%%%%
In this section, we present an application of the presented automatic procedure to a real database. The images have been obtained by the Department of Basic Health Sciences, Faculty of Health Sciences of the URJC in Madrid.

{
\subsubsection{Description of the dataset and data collection} }

 Immunoreactivity of spinal cord sections (L3-L5) was performed for ionized calcium-binding adapter molecule 1 (Iba-1) as a marker of microglia using diaminobenzidine (DAB) immunohistochemistry.\footnote{{DAB is most often used in immunohistochemical staining as a
chromogen. It is oxidized in the presence of peroxidase and hydrogen peroxide in the standard
immunohistochemistry protocol with HRP-DAB detection. This results in a typically brownish
color that, in case of Iba-1 immunohistochemistry (a microglia/macrophage-specific cytoplasmic
calcium-binding protein) stains all the cytoplasm of microglial cells, making it easy to identify
their shapes.}} 
{
Tissue processing and immunohistochemistry were performed as following (and described in \cite{GARCIA2022112986}): briefly, animals were deeply anesthetized with sodium pentobarbital ( $60 \mathrm{mg} / \mathrm{kg}$, i.p.) and intracardially perfused via the left ventricle with 200 ml of normal saline followed by fresh-prepared $4 \%$ paraformaldehyde (PFA) in 0.1 M phosphate buffer saline (PBS), $\mathrm{pH} 7.4,4 \mathrm{C}$. A laminectomy of the entire thoracic and lumbar spinal cord was performed, and a L3-L5 spinal segment was excised and post-fixed in the same fixative solution for four extra hours at room temperature. Then cryopreserved in $0.1 \mathrm{M} \mathrm{PBS}(\mathrm{pH} 7.4)$ supplemented with $30 \%$ sucrose and $0.05 \%$ azide at 4C until further processing.}
{ Following, post-fixation spinal cord specimens were included in paraffin. Paraffin wax blocks were thaw mounted in a Minot rotatory microtome and serial 5 m transverse sections were collected on $0.02 \%$ poly-L-Lysine coated slides. After deparaffinization, samples  followed microwave antigen retrieval in citrate buffer. Then treated with fresh PBS containing $3 \%$ hydrogen peroxide for 15 min to block the activity of endogenous peroxidase, followed by 30 min incubation with bovine serum albumin (BSA) to reduce nonspecific background staining. Immunodetection was performed with goat anti-Iba-1 (1:500, Abcam5076) overnight, 4C. For secondary detection, the specimens were incubated for 1 hour with CyTM2-conjugated donkey anti-goat serum immunoglobulin (1:350, Jackson ImmunoResearch). Visualization was achieved using a freshly prepared solution of $0.05 \%$ DAB (Sigma-Aldrich) for 5 min . Eventually, sections were counterstained with Harris hematoxylin, washed well in tap water, dehydrated, and coverslipped with EukittTM mounting medium (Sigma-Aldrich). The resulting specimens were observed under a Zeiss Axioskop 2 microscope equipped with the image analysis software package AxioVision 4.6, and series of photomicrographs were recorded to make montages of entire spinal cords at final magnifications of 20x by means of the Microsoft Image Composite Editor (ICE) software.
}
 The immunohistochemistry technique yields a dark brown color reaction. However, heat induced epitope retrieval (HIER), deficient rising/washing, step or excessively high concentrations of DAB or hematoxylin counterstain can ultimately cause burgundy, blackish, brownish or dark purple staining artifacts.
 {True positive assessment was also sustained on manual cell counting. In the dataset images, a green line was drawn to separate RexedÕs lamina VI from the ventral horn according to an anatomical atlas for lumbar sections L3, L4 or L5. We have studied and analyzed only the superior part of the images (above the green line).}
 Thus, $D=12$ images were processed considering the expert opinion (i.e., manual counting). Anything other than the grey matter was removed, and the background was colored in white.

\subsubsection{Results}
\label{ResSubsect}
We apply the presented technique to the stored database of $D$ images. We perform a LOO-CV procedure minimizing the $L_{\infty}$ distance (i.e., minimizing the maximum error) between the predicted number and the true number of microglial cells. We obtain an $L_{\infty}$  error curve as a function of $1/\eta$, where there is a flat zone of minimum values between $1/\eta\in [1.3, 3.75]$. We choose an intermediate value setting $1/\eta^*\approx 2.5$, i.e., $\eta^* \approx 0.40$.  With the values of $\eta$ such that $1/\eta\in [1.3, 3.75]$, the maximum error in prediction never exceeds $25$ units. Just to show the robustness of the technique, we also provide the results of LOO-CV procedure minimizing the $L_1$ distance:
$1/\eta*\approx 0.85$, i.e., $\eta^* \approx 1.17$. With this value of $\eta$, we obtain an averaged absolute error in  LOO-CV procedure of less than 10 units in prediction. Considering all the data points and $\eta^* \approx 0.40$, 
the average absolute error decreases to $3.68$, i.e., less than $4$ units. 
 Note that, we are obtaining already remarkable results with a small and heterogenous dataset. 
\newline
 In Figure \ref{FigMSEvTotro2}, we depict the points with the numbers of cells given by the expert as $x$-values and the numbers of cells  predicted  by the KC method  as $y$-values, using $\eta^* \approx 0.40$ obtained with LOO-CV above. Ideally, we would like to have all the points belonging to the black straight line $x=y$ in Figure \ref{FigMSEvTotro2}. This ideal case would correspond to the perfect prediction of all numbers of microglial cells. We can observe that the points are all very close to the black straight line  with a coefficient of determination of $R^2\approx 0.90$ (in an ideal scenario, we would have $R^2=1$). Considering all the data points (instead of leave-one-out as in LOO-CV), we obtain an average absolute error less than $4$ units with a standard deviation of $5.44$.
 \newline
 Moreover, the KC algorithm also provides uncertainty estimations of its predictions, i.e., $\widehat{\sigma}_d^2$: the error bars show $\pm 2\sqrt{\widehat{\sigma}_d^2}$, corresponding to the $95\%$ of the probability. Note that the error bar always contains the black line; hence, the expert's opinion is always contained in the uncertainty interval.  Moreover, the three points virtually belong to the black straight line, having (all of them) uncertainty intervals with almost zero length. This shows that the algorithm is working extremely well estimating properly the uncertainty.
The points with higher uncertainty can suggest a need for the revision of the image by the expert.\footnote{Related Matlab code is given at: \url{http://www.lucamartino.altervista.org/PUBLIC_CODE_KC_microglia_2025.zip}. }
\newline
\newline
{{\bf Comparisons.}
We compare the obtained results with different techniques in the literature, in a  dataset with 12 images. Firstly, we compare with the method proposed in \cite{khakpour2022manual} that uses  the software ImageJ. We also apply two convolutional neural networks (CNNs) proposed in the literature in \cite{anwer2023comparison}, with the code available at \url{https://gitlab.com/cell-quantifications/Microglia}. 
The results are shown in Table \ref{tab:tablecomp}. We made a special effort to obtain the best possible performance for the alternative procedures (testing several values of the parameters etc.). However, the proposed KC scheme returns the best results. We believe that especially the CNNs (traning over the high-resolution images) require a much bigger number images in the  dataset.

% {\color{red}
% \textbf{Outline of algorithm in \cite{khakpour2022manual}}
% The automation process presented in \cite{khakpour2022manual} consist on a series of steps in software ImageJ performed sequentially. After completing the process they are able to ¿locate the microglial cells? Between the main steps of this algorithm we have the:
% \begin{enumerate}
%     \item Scaling and converting to grayscale the image.
%     \item Curation of the image by enhancing the contrast of the image and  automatic adjustment of the threshold using maximum entropy (AQUÍ NO EXPLICAN SI ESTO ES UN COMANDO QUE YA VIENE DEFININDO EN IMAGEJ), removing peakles and outliers.
%     \item Finally, analyze the image to detect microglial cells with size between predefined limits (DICEN QUE EJECUTAN EL COMANDO ``ANALYSE PARTICLES'', DE NUEVO, NO SÉ SI ES UN COMANDO PROPIO O VIENE EN IMAGEJ
% \end{enumerate}

\begin{table}[h!]
{
  \caption{Results in terms of $R^2$ with different schemes.}
\label{tab:tablecomp}
 \centering
 \begin{center}
\begin{tabular}{c|c|c|c|c|}
  \cline{2-5}
\multicolumn{1}{c|}{} & {\bf KC} & {\bf Method in \cite{khakpour2022manual} } & {\bf CNN-1 \cite{anwer2023comparison}} &  {\bf CNN-2 \cite{anwer2023comparison}}\\
  \cline{2-5}
   \hline
  \multicolumn{1}{|c|}{$R^2$} & 0.90 & 0.67 & 0.70 & 0.74 \\
   \hline
%  time & 100\% & 150\% & 3967\% & 5530\%  \\
 % \hline
\end{tabular}
\end{center}
}
\end{table}

\noindent
Additionally, we have applied the proposed KC, with a dataset where we apply  data augmentation as described in Section \ref{dataAug}, to an image provided in the GitLab repository of \cite{anwer2023comparison}. Testing their models with this provided image they obtain as prediction 97 microglial cells. Applying the proposed KC to the same image, we obtain 95 with a very small variance. This is an additional confirmation of the good performance provided by the proposed algorithm.  %The corresponding image is also given at \url{ http://www.lucamartino.altervista.org/PUBLIC_CODE_KC_microglia_
%2025.zip}.
}

\section{Conclusions}\label{SectionConc}
We have proposed an automatic counter of microglial cells in immunohistochemical images. Automated cell counting is superior to manual cell counting, at least in terms of  human resource  efficiency, and the proposed scheme provides similar accuracy.  Conventional manual cell counting is dependent on the researcher's expertise and is time-consuming. 
The proposed KC algorithm is a counter; indeed, the predictions/estimations are always non-negative, and the predictions can easily be converted into integers by rounding them. 
{The proposed scheme simplifies the database creation by
reducing the manual effort required from human operators: (a) the pixel cell annotations is  not required (but only the total
number of cells within each analyzed image region); (b) the images can have different size so that, only portions of images can be included, which further simplifies the expert’s labeling task, as different regions of the same image can be incorporated at different stages.}
KC also provides an uncertainty  estimation of the predictions (e.g., see  Eq. \eqref{VAR_PRED_form}). In the smoothing case, if a smoothed value presents a high uncertainty, this can indicate a need for revision of the image by the expert.
Furthermore, the multi-experts scenario can be directly handled by the novel scheme.
\newline 
{The proposed scheme is formed by two main components. The first part (P1) is a tailored feature extraction that  is applied to suppress irrelevant components within the high-dimensional images, enhancing the quality of the input by boosting the signal-to-noise ratio. The second part (P2) is a regression scheme.} 
The designed regression method has a unique hyperparameter to learn, that is the non-negative scalar $\eta$, which can be learnt by leave-one-out cross-validation (LOO-CV). For this reason, the proposed KC is easy and fast to train even in small datasets.   Even if we have only one hyperparameter to learn, the method is a {\it non-parametric} regressor, i.e., the  complexity of the solution in Eq. \eqref{PRED_form} grows with $D$ (that is,  the number of images in the database). Moreover, the solution is {\it non-linear}  with respect to the inputs. Hence, the proposed method is able to express the complexity of rich datasets.  However, other possible KC versions (with more hyperparameters) have also been discussed.
 {It is important to remark that a practitioner may substitute the kernel smoother in P2 with a deep learning method (or any alternative), if it meets all the requirements outlined in Section \ref{KeySect}.}
 Last but not least, the results obtained in the different experiments are very promising. Related Matlab code has been also provided in order to facilitate the use by interested practitioners.
\newline
{ Theoretical aspects have also been discussed. For instance, a related statistical model has also been provided in App. \ref{STATmod} to study in synthetic experiments the behavior of the proposed KC algorithm. A detailed discussion on the optimal choice of the threshold vectors has been given in App. \ref{SectionChoice}. }
Finally, note also that  the problem of counting objects in images (even different from microglial cells) is still one of the relevant tasks in different applications: for instance, counting cells in microscopic and biomedical images, monitoring crowds in surveillance systems, counting the number of green spots in satellite images  \cite{ZHANG2024121602,paul2021object,rahman2013counting} etc.
Therefore, the proposed approach can have a vast range of applications, since it can be employed for counting different objects in different types of images, for instance, provided  by a  satellite, telescope, or drones, to name a few
\citep{konatar2020box,lempitsky2010learning,nasor2020detection,SVENDSEN2020107103,LLORENTE21DSP}.

%We have introduced an automatic counting algorithm for computing the number of microglial cells in biomedical images.  We have designed a linear predictor based on the information provided by filtered images, obtained applying color threshold values to the labelled images in the dataset. Furthermore, in order to reduce the sensitivity to the threshold value, filter type, lighting conditions,  the chosen thresholds are converted into quantile values.
% Non-linear extensions and other improvements, jointly with the choice of the threshold values, have  been discussed. Different numerical experiments show the capability and consistency of the proposed algorithms, even with noisy labelled images.
%\newline
%The novel approach allows to increase its flexibility and its performance in a very simple and automatic way: considering and adding more threshold vectors and, as a consequence, more columns in the design matrix. Clearly, the proposed schemes could be directly applied to different counting problems of small objects in other types of images.
%As future lines, we also plan the use of ANNs and the total least squares method due to the presence of noise in the labels (or even more complex techniques as in \citep{Morelli21}). Increase the database by artificial images requires additional studies as well.

%\section{some columns can depend also on quantiles...}

%\section{use the possible correlation among the $3$ outputs}

%\section{inducing sparsity in $\beta$ can help also the detection...in some way....``choose some point''}

\section*{Acknowledgement}

 The work was partially supported by Agencia Estatal de Investigaci{\'o}n AEI, Spain with project SP-GRAPH, ref. num. PID2019- 105032GB-I00, and project POLI-GRAPH, Grant PID2022-136887NB- I00, by project Starting Grant for Rttb, BA-GRAPH ÔÔEfficient Bayesian inference for graph-supported dataÕÕ, of the University of Catania (UPB-28722052144),  by the project LikeFree-BA-GRAPH funded by ÔÔPIAno di inCEntivi per la RIcerca di Ateneo 2024/2026ÕÕ of the University of Catania, Italy, and by
Research and Development (R\&D) project for young PhD graduates in the Region of Madrid (CAM)-URJC, F861-3483, years 2022-2024 to L.M. P.P. was awarded with a collaboration scholarship from the Spanish Ministry of Education and Vocational Training (call 2021-2022).

%\bibliographystyle{elsarticle-num} 
%\bibliographystyle{IEEEtran}
%\bibliography{bibliografia,biblioFading}
%\bibliography{bib_battery,other,traffic_new,bibliografia}
\bibliographystyle{plain}
\bibliography{bibliografia}

%%%%%%%%
\appendix
%%%%%%%%
%%%%%%%%%%%%%%%%%%%%%%%%%%%%%%%%%%%%%%%
%\section{Appendix 1}
%\label{App1}
%%%%%%%%%%%%%%%%%%%%%%%%%%%%%%%%%%%%%%%

%\newpage

%%%%%%%%%%%%%%%%%%%%%%%%%%%%%%%%%%%
\section{Related statistical model}\label{STATmod}
%%%%%%%%%%%%%%%%%%%%%%%%%%%%%%%%%%%

\begin{figure}[h!]
    \centering
       \centerline{
   \includegraphics[width=18cm]{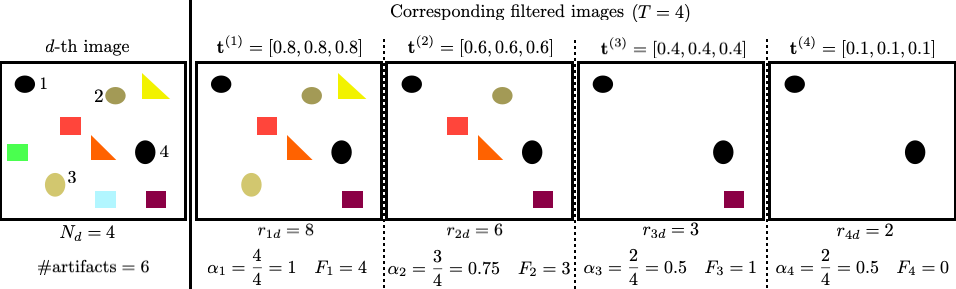}
    }
    \vspace{-0.3cm}
    \caption{{\footnotesize We recover the illustrative example in a previous figure: a generic $d$-th image is filtered $T=4$ times with different threshold vectors ${\bf t}^{(k)}$, obtaining the  corresponding filtered image. In this example, we have $N_d=4$ microglial cells. In each filtered images, we also show the total number of objects $r_{kd}$, the percentage $\alpha_k\in[0,1]$ of the contained microglial cells, and the number $F_k$ of other objects (artifacts).   } }
    \label{FigSUPER_resumen}
\end{figure}

In this section, we provide a statistical model that can be employed for synthetic experimental analysis and study the theoretical behavior of the proposed KC algorithm.
Each filtered image will contain a certain percentage $\alpha\in[0,1]$ of the total  number of microglial cells $N_d$ ( true positives, $\lfloor\alpha N_d \rceil=\#TP$) and also a certain number $F$ of other objects ( false positives $F=\#FP$, or number of artifacts in the filtered image) that do not represent microglial cells. 
\newline
\newline
Mathematically speaking, we can express the behavior above with the following statistical model,
 \begin{align}\label{EqModel}
r_{kd}&=\lfloor  \alpha_k N_d+F_k \rceil, \quad \alpha_k\in (0,1], \quad k=1,...,T, \mbox{ }\mbox{ }d=1,...,D, \\
F_k &\in\{v_k,v_k+1, v_k+2,...,s_k\} \quad s_k,v_k\in \mathbb{N}^+, \quad s_k>v_k,
 \end{align}
where $\lfloor b \rceil$ returns the nearest integer to $b$; the coefficient $\alpha_k\in (0,1]$  depends on the ${\bf t}^{(k)}$, i.e., $\alpha_i=\alpha({\bf t}^{(k)})$, and $F_k$ is a discrete random variable which takes non-negative integer values, contained in  $[v_k, s_k]$.
%$F\in\mathbb{N}=\{0,1,2,3,...\}$.
 The other coefficients $s_k,v_k\in \mathbb{R}^+$, are positive real values depending also on ${\bf t}^{(k)}$, i.e., $s_k=s({\bf t}^{(k)})$, and $v_k=v({\bf t}^{(k)})$. Clearly, $s_k$ affects the support of the random variable $F_k$ such that
 \begin{align}
v_k&= \min F_k \in  \mathbb{N}^+,  \\
 s_k&=\max F_k \in  \mathbb{N}^+. 
% \mbox{namely, } F_i &\in [v_i, s_i]  \nonumber
 \end{align}  
We have $v_k< \infty$ whereas $s_k$ could be infinity as well. The probability mass function (pmf) of $F_k$ defined in this support can change depending on the specific application. Note that, with this model, we are assuming that $\alpha_k$, $s_k$, $v_k$ and $F_k$ are in some sense stationary quantities/variables, which do not depend on the specific image to analyze. However,  $\alpha_k=\alpha({\bf t}^{(k)})$, $v_k=v({\bf t}^{(k)})$ and $s_k=s({\bf t}^{(k)})$ depend on the threshold vector, so that the {\it signal-to-noise ratio} (SNR) is different in each filtered image. Note that $\#TP=\lfloor\alpha_k N_d \rceil$ and $\#FP=F_k$.
\newline
 Note that Eq. \eqref{EqModel} shows the relationship between each $r_{kd}$ and $N_d$. Since $\alpha_k\in (0,1]$ and it multiplies the number of microglial cells $N_d$ in the $d$-th analyzed image, the value $100 \alpha_k \%$ is the percentage of microglial cells in the $d$-th filtered image (i.e, percentage of true positives), whereas $F_k$ is the number of  false positives. Thus, $s_k$ and $v_k$ are the maximum and minimum possible number of false alarms in the image filtered with the threshold vector ${\bf t}^{(k)}$. Finally, since $\alpha_k=\alpha({\bf t}^{(k)})$, $s_k=s({\bf t}^{(k)})$ and $v_k=v({\bf t}^{(k)})$, we can also write
  \begin{align}\label{EqNecparaSIMU}
  v_k=v(\alpha_k), \quad s_k=s(\alpha_k),
 \end{align}
 i.e., $v_k$ and $s_k$ can also be seen as functions of the percentage of the number of microglial cells in the $k$-th filtered image. %A summary of the main notation is given in the Table \ref{TableNotation}.
  An illustrative example of filtered images and the corresponding $r_{kd}$ values is given in Figure  \ref{FigSUPER_resumen}. Assuming this model, an upper bound and a lower bound for $N_d$ can be obtained as shown in Appendix \ref{MessaAdessoApp}.

 %%%%%%%%%%%%%%%%%%%%%%%%%%%%%%%%%%%%%%%%
\section{Choice of the threshold vectors }\label{SectionChoice}
%%%%%%%%%%%%%%%%%%%%%%%%%%%%%%%%%%%%%%%%
%Given the 

The proposed algorithm works for any possible choice of ${\bf t}^{(k)}$, with the unique requirement that the vectors ${\bf t}^{(k)}$, with $k=1,...,T$, must be different to each others.
However, the performance of the algorithm depends on the choice of threshold vectors, 
$$
{\bf t}^{(k)}\in [t_1^{(k)},t_2^{(k)}, t_3^{(k)}] \in [0,1]^3,
$$
 with $k=1,...,T$, that are employed  to obtain $T$ different filtered images from each  $d$-th image in the database.
%However, it is also important to remark that, as $T$ grows and as $T \rightarrow D$,  the dependence of the performance on the choice of ${\bf t}^{(k)}$ tends to decrease.
This is actually an experimental design and/or active learning problem: see, for instance, \cite{BUSBY20091183,LLORENTE21DSP,SVENDSEN2020107103}.
 In this section, we discuss some required concepts and a procedure for a proper choice of ${\bf t}^{(k)}$.
\newline
\newline
Note that large values of $t_i^{(k)}$ (close to 1) ensure the presence of microglial cells, but also a huge number of other artifacts that are false alarms. Small values of $t_i^{(k)}$ (close to 0) ensure to substantially reduce the number of false alarm; however,  some microglial cells can be missed. Considering the model in Eq. \eqref{EqModel}, we have that $\alpha_k=\alpha({\bf t}^{(k)})$ and $s_k=s({\bf t}^{(k)})$; hence, the {\it signal-to-noise ratio} (SNR) is different in each filtered image (different $\#TP$ and $\#FP$).  For our purpose, the SNR can be defined as 
$$
\mbox{SNR}=\frac{\#TP}{\#FP}=\frac{\alpha N_d}{F}.
$$
Clearly, the vectors ${\bf t}^{(k)}$ corresponding to larger SNR values are preferable. 
{\rem\label{RemImpOptVect} The idea is to choose the vectors ${\bf t}^{(k)}$ that provide the higher values of $\mbox{SNR}=\frac{\#TP}{\#FP}$. For instance, this can be done directly by a Monte Carlo search. However, below we divide this search into two steps.}
\newline
\newline
Another interesting observation is that different vectors ${\bf t}^{(k)}$ can provide the same number of true positives $\#TP$, but a different number of $\#FP$. Clearly, given a value $\#TP$ , we prefer the vectors ${\bf t}^{(k)}$ that provide the smallest value of false positives $\#FP$. We call them {\it optimal conditional thresholds}.
\newline
Therefore, conceptually the search of proper threshold vectors can be divided in two steps:
\begin{itemize}
\item First of all, fixing a value of true positives $\#TP$, find the optimal conditional thresholds, i.e., the thresholds that maximize the SNR, providing the smallest number of false positives $\#FP$.  
\item Among the optimal conditional thresholds previously obtained, choose the thresholds corresponding to the largest SNRs (possibly $\mbox{SNR}>1$).
\end{itemize}
 With the first step, we can build the curve ``minimum number of $\#FP$'' versus $\#TP$ (or percentage of $\#TP$), as shown in Figures \ref{CondOptimalFig} and \ref{CondOptimalFig2}. This allows to detect the range of $\#TP$ values with SNR$>1$, as depicted in Figure  \ref{CondOptimalFig_otra}. Moreover, the division in these two phases, also allows the use of a different  payoff function in the second step, instead of maximizing the SNR (i.e., in the proposed procedure, the payoff function is the SNR).
 %%%%Recall each point of the curve in Figure \ref{CondOptimalFig} correspondien
The following two subsections describe these two steps.

%{\rem As highlighted in Remark \ref{RemImpOptVect}, there is no need of dividing the problem in two steps. We can directly look for  the vectors ${\bf t}^{(i)}$ which provide the higher values of $\mbox{SNR}$. However, the concept division above allows }

%%%%%%%%%%%%%%%%%%%%%%%%
\subsection{Optimal conditional thresholds}
\begin{figure}[h!]
    \centering
    \centerline{
  %  \subfigure[\label{CondOptimalFig}]{\includegraphics[width=8cm]{NumberObjectsDetectionPercentage_v2.png}}
  %  \subfigure[\label{CondOptimalFig_otra}]{\includegraphics[width=8cm]{NumberObjectsDetectionPercentage_v2.png}}
   \includegraphics[width=8cm]{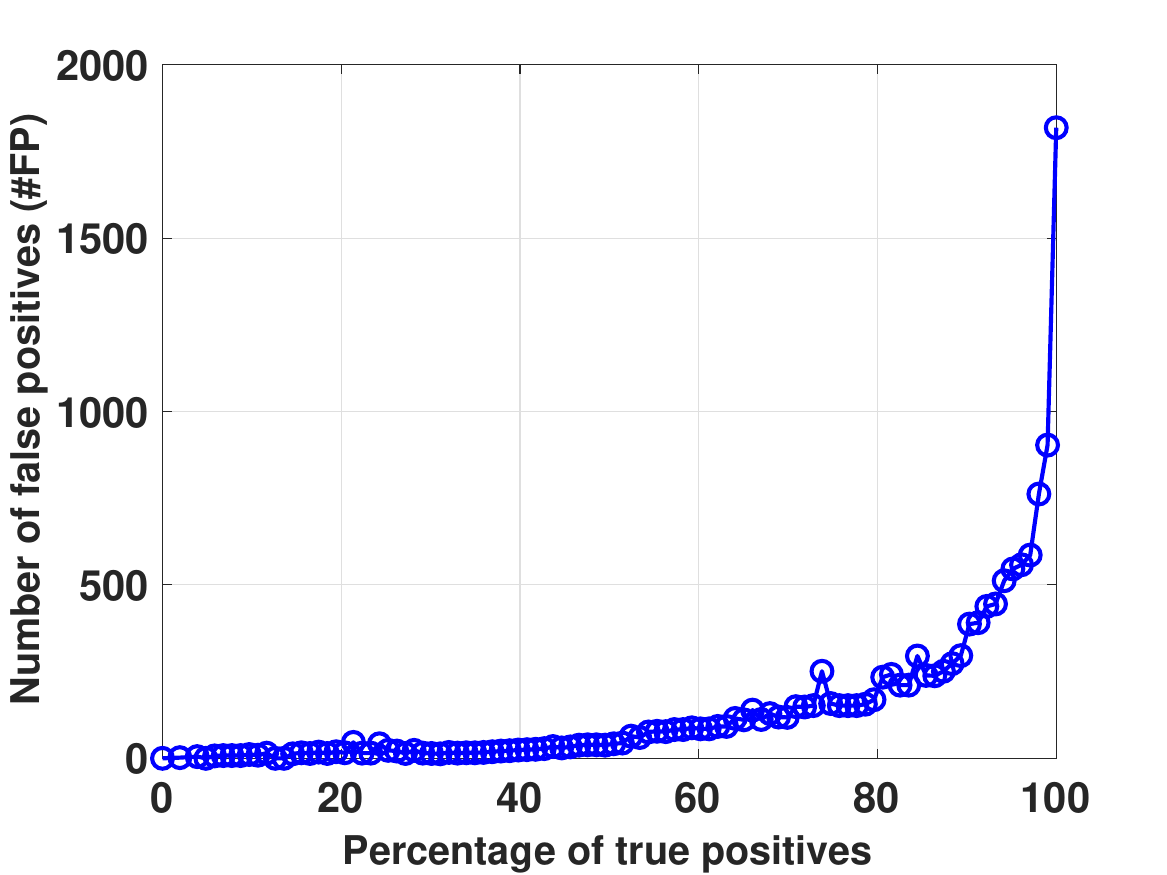}
    }
    \caption{{\footnotesize Example of minimum possible number of possible false alarms (false positives) ($\#FP$) versus the percentage of true positives ($\frac{\#TP}{N_d}100\%$) in a specific image of the dataset, obtained by the corresponding optimal conditional thresholds. By increasing the values of thresholds, more $\#TP$ but also more $\#FP$ are obtained. Here, the minimum possible number of  false alarms for each given value of $\#TP$ is shown. } }
    \label{CondOptimalFig}
  %    \label{CondOptimalFig_all}
\end{figure}

%%%%%%%%%%%%%%%%%%%%%%%%
First of all, we have to notice that, given a threshold vector ${\bf t}\in \mathbb{R}^3$, in the filtered image we have a certain number of true positives $\#TP$ (microglial cells), and a certain number of  false positives $\#FP$ (i.e., false alarms). Two different threshold vectors ${\bf t}_1, {\bf t}_2$ could give the same number of true positives $\#TP$ but different values of false positives $\#FP$. Clearly, given a number of true positives $\#TP$, we prefer the threshold vector that provides the smallest number of false positives (i.e., that minimizes $\#FP$). We call this vector as {\it optimal conditional  threshold vector}, i.e., the optimal  vector corresponding to the  true positive value $\#TP$. Namely, the optimal conditional  vector maximizes the signal-to-noise ratio (SNR), fixing $\#TP$,  
$\mbox{SNR}=\frac{\#TP}{\#FP}$.
In order to find the optimal conditional  threshold vectors, we employ a Monte Carlo search:
\newline
\newline
%Given a value of $\#TP$ (or a percentages of true positives):
\fbox{
\parbox{\textwidth}{
\begin{enumerate}
\item  For $s=1,...,M_{\texttt{runs}}$:
\begin{itemize}
\item Draw $t_{i,s}\sim \mathcal{U}([0,1])$, for $i=1,2,3$.
\item Set  ${\bf t}_s=[t_{1,s},t_{2,s},t_{3,s}]$ and obtain the filtered image corresponding to ${\bf t}_s$.
\item Count the number of true positives $\#TP(s)$, and false positives $\#FP(s)$  in this filtered image.
\end{itemize}
\item For each value $\gamma=1,2...,N_d$:
\begin{itemize}
\item Find all the indices $s^*$ such that $\#TP(s^*)=\gamma$, defining a set of indices $\mathcal{S}_\gamma$.
\item For all $s^*\in\mathcal{S}_\gamma$:
\begin{itemize}
\item Find 
\begin{eqnarray}
 s^{\texttt{opt}}_{\gamma}=\arg\min_{s^*\in\mathcal{S}_\gamma} \mbox{ }\#FP(s^*).
\end{eqnarray}
\item Then the optimal conditional vector  of thresholds (for $r$ true positives) is ${\bf t}_{s^{\texttt{opt}}_{\gamma}}=\left[t_{1,s^{\texttt{opt}}_{\gamma}},t_{2,s^{\texttt{opt}}_{\gamma}},t_{3,s^{\texttt{opt}}_{\gamma}}\right]$.
\end{itemize}
\end{itemize}
\end{enumerate}}}
\newline
\newline
Therefore, given a pre-established number of true positives $ \gamma=\#TP$ with $1\leq \gamma \leq N_d$, the optimal conditional  threshold vector is  ${\bf t}_{s^{\texttt{opt}}_{\gamma}}$: this is the vector that provides the minimum possible number of false alarms $\#FP$ (given the pre-established value $\#TP=\gamma$). Clearly, ${\bf t}_{s^{\texttt{opt}}_{\gamma}}$ is an estimation of the optimal vector due to the Monte Carlo procedure: as $M_{\texttt{runs}}\rightarrow \infty$, this approximation improves. {This procedure should be repeated for as many images as possible in the training set.}
\newline
In our specific application, we use $M_{\texttt{runs}}=10^4$ independent runs. Figure \ref{CondOptimalFig} depicts the  minimum possible number of false alarm objects ($\#FP$), obtained by the optimal conditional  thresholds, versus the percentage of true positives ($\frac{\#TP}{N_d}100\%$). Clearly, there is a trade-off between the number of false alarms and true positives. %Both values are directly correlated to each other, indeed, it may not be possible to detect all cells in the figure without dramatically rising the number of false alarms.

% In Table \ref{optimalthresholds},
% the {\it Optimal conditional  thresholds} are shown,  given a fixed number of true positives, i.e., keeping fixed $\#TP$, and by looking inside the $S=10^4$ independent runs of random search that were performed. Namely, given a number of true positives $\#TP$, searching for the $s_*$-th run that minimizes the number of false positives $\#FP$ (i.e., false alarms) with the choice ${\bf t}^{(s_*)}=[t_1^{(s_*)},t_2^{(s_*)},t_3^{(s_*)}]$.

%%%%%%%%%%%%%%%%%%%%%%%%%%%%%%%%%%%%%%%%%%%%%
\subsection{Choosing among the optimal conditional thresholds}
%%%%%%%%%%%%%%%%%%%%%%%%%%%%%%%%%%%%%%%%%%%%%

\begin{figure}[h!]
    \centering
    \centerline{
    \subfigure[\label{CondOptimalFig2}]{\includegraphics[width=8cm]{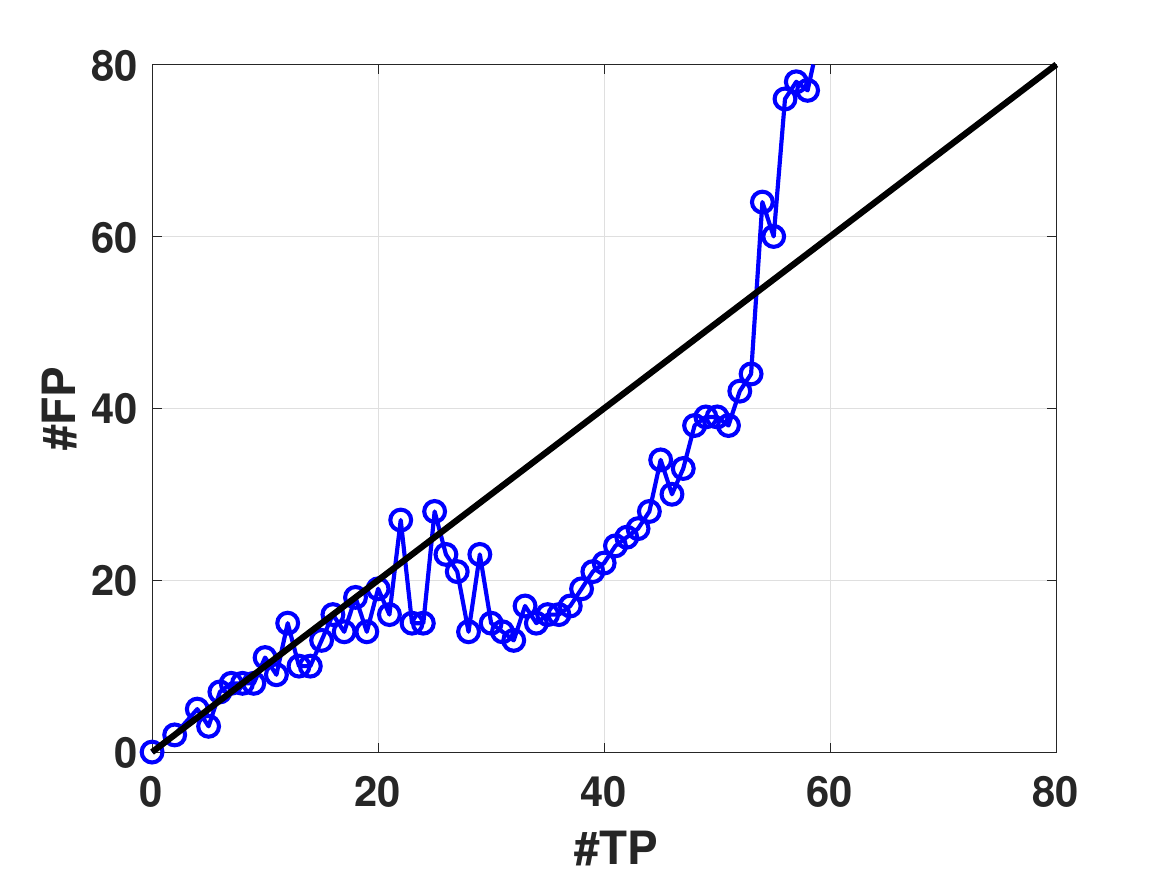}}
    \subfigure[\label{CondOptimalFig_otra}]{\includegraphics[width=8cm]{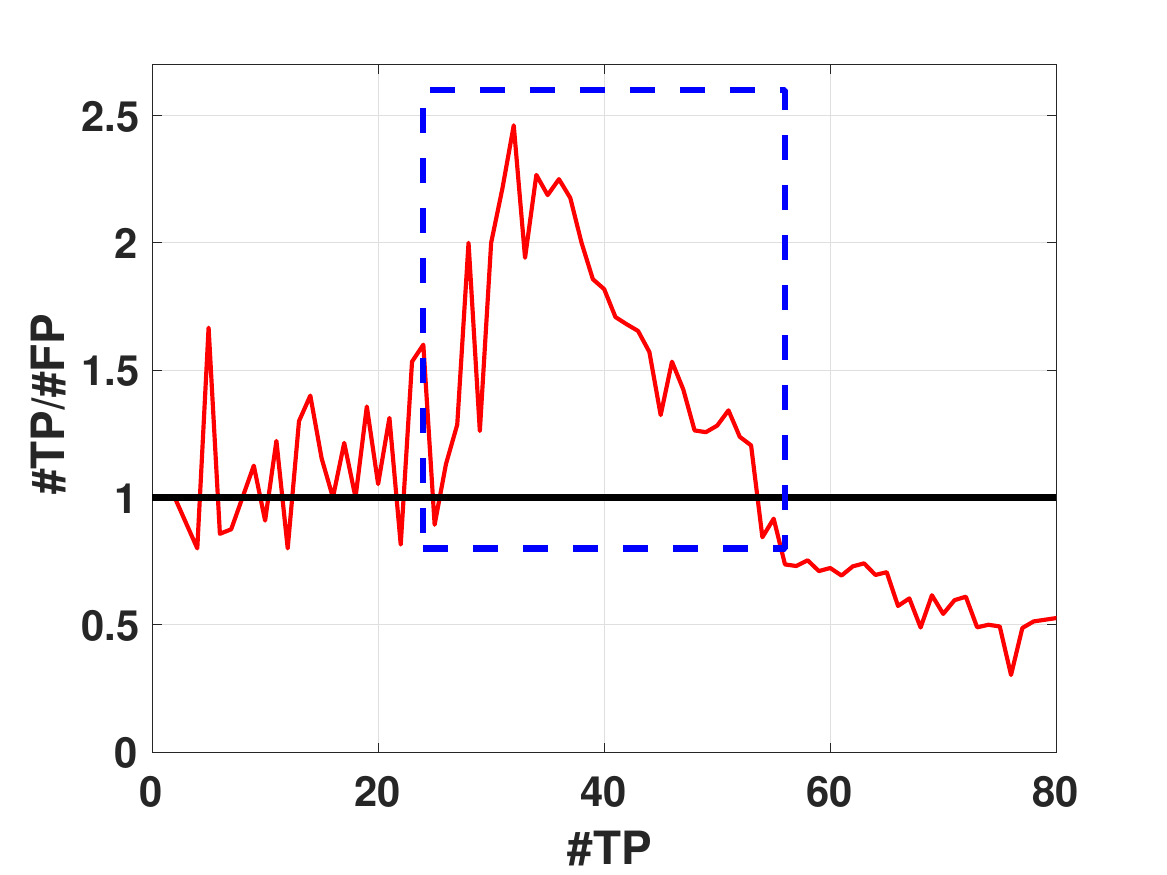}}
    }
    \caption{{\footnotesize {\bf (a)} The minimum number of false positives ($\#FP$) obtained by the optimal conditional thresholds, versus the number of true positives ($\#TP$), in a range of values where we can obtain $\mbox{SNR}\geq1$ (for a specific image of the dataset). The solid black line represents the straight line $\#FP=\#TP$, then when the blue curve is below the black straight line, we have $\mbox{SNR}\geq1$. {\bf (b)} The $SNR=\frac{\#FP}{\#TP}$ corresponding to the values in Figure \ref{CondOptimalFig}. The rectangular region depicted by dashed lines, shows the values of SNR bigger than 1 (obtained by the optimal conditional thresholds for $25\leqÊ\#TP\leq 53$).   } }
    \label{FigSNR}
\end{figure}

Each point of the curve in Figure \ref{CondOptimalFig} corresponds to an optimal conditional vector ${\bf t}_{s^{\texttt{opt}}_{\gamma}}$  with $1\leq \gamma \leq N_d$ and $\#TP=\gamma$. Figure \ref{CondOptimalFig2}, shows the same curve in  Figure \ref{CondOptimalFig} but, in this case, the values of $\#TP$ are directly given on the horizontal axis. Moreover, the straight line $\#FP=\#TP$ is depicted with a solid black line.  Clearly,  when the blue curve is below the black straight line, we have 
$$
\mbox{SNR}=\frac{\#TP}{\#FP}>1.
$$
The corresponding values of the SNR are provided in Figure  \ref{CondOptimalFig_otra}. The rectangular region depicted by dashed lines shows the values of SNR bigger than 1 (obtained by the optimal conditional thresholds for $25\leqÊ\#TP\leq 53$). Hence, we should choose $T$ points in this interval, e.g., in this example (corresponding to results from our database), $ 25\leqÊ\gamma_k\leq 53$, $i=1,...,T$,
  and we use the corresponding optimal conditional vector ${\bf t}_{s^{\texttt{opt}}_{\gamma_k}}$.

%%%%%%%%%%%%%%%%%%%%%%%%%%%%%%%%%
\section{Lower and upper bounds}\label{MessaAdessoApp}
%%%%%%%%%%%%%%%%%%%%%%%%%%%%%%%%%

Let us consider both $v_i< \infty$, $s_i<\infty$, finite and known {in the model given in App. \ref{STATmod} (recall also that $0\leq v_i<s_i$).} Moreover, let us also assume the coefficients $\alpha_i\in (0,1]$, as known values (or approximately known - estimated). In this scenario, if the observations $r_{id}$ are generated according to  the model 
 \begin{align*}
r_{id}&=\lfloor  \alpha_i N_d+F_i \rceil, \quad \alpha_i\in (0,1], \quad i=1,...,T, \mbox{ }\mbox{ }d=1,...,D,
%F_i &\in\{v_i,v_i+1,...,s_i\} \quad s_i,v_i\in \mathbb{N}^+, \quad s_i>v_i,
 \end{align*}
we can obtain some lower and upper bounds {for $N_d$. Indeed,} since $F_i\in [v_i,s_i]$, we have
 \begin{align}
%r_{id}&=\lfloor  \alpha_i \widehat{N}_{d,\texttt{upper}}+v_i \rceil  \longrightarrow 
 \widehat{N}_{d,\texttt{lower}}=\left\lfloor \frac{r_{id}-s_i}{\alpha_i} \right\rfloor, \qquad
\widehat{N}_{d,\texttt{upper}}=\left\lceil \frac{r_{id}-v_i}{\alpha_i} \right\rceil.
%r_{id}&=\lfloor  \alpha_i \widehat{N}_{d,\texttt{lower}}+s_i \rceil 
 \end{align}
where $\lfloor b \rfloor$ gives as output the greatest integer less or equal to $b$, whereas  $\lfloor b \rfloor$ returns  the least integer greater or equal to $b$. Namely,  if the observations $r_{id}$ are generated according to the model above,  we can assert that the true number $N_d$ of microglial cells is contained (with probability $1$) in the following interval: %our estimator $ \widehat{N}_d$ should satisfy
  \begin{align}
 \widehat{N}_{d,\texttt{lower}} \leq  N_d \leq\widehat{N}_{d,\texttt{upper}}.
 \end{align}
Therefore, by estimating $s_i$ and $v_i$ (for a certain percentage $\alpha_i$ of microglial cells in the filtered images) we can obtain lower and upper bounds for $N_d$, using the inequalities above.

\end{document}